\documentclass[10pt,superscriptaddress,aps,prc,twocolumn]{revtex4-1}

\usepackage[dvipsnames]{xcolor}
\usepackage{braket}
\usepackage{cancel}
\usepackage{fancyvrb}
\usepackage{tabularx} 
\usepackage{graphicx} 
\usepackage{hyperref}  
\usepackage{amssymb}  
\usepackage{amsmath}  
\usepackage[]{units}

\begin{document}
\title{Bias-Variance Trade-off and Model Selection for Proton Radius Extractions}

\author{Douglas~W.~Higinbotham}
\affiliation{Jefferson Lab, Newport News, VA 23606}

\author{Pablo Giuliani}
\affiliation{Department of Physics, Florida State University, Tallahassee, Florida 32306, USA}

\author{Randall~E.~McClellan}
\affiliation{Jefferson Lab, Newport News, VA 23606}


\author{Simon~\v{S}irca}
\affiliation{Faculty of Mathematics and Physics, University of Ljubljana, SI-1000 Ljubljana, Slovenia}
\affiliation{Jo\v{z}ef Stefan Institute, SI-1000 Ljubljana, Slovenia}

\author{Xuefei~Yan}
\affiliation{Duke University, Durham, NC 27708}

\begin{abstract}
Intuitively, a scientist might assume that a more complex regression model will necessarily yield a 
better predictive model of experimental data.
Herein, we disprove this notion in the context of extracting the proton charge radius from charge 
form factor data. Using a Monte Carlo study, 
we show that a simpler regression model can in certain cases be the better predictive model. 
This is especially true with noisy data where the complex model will fit the noise instead of the physical signal. 
Thus, in order to select the appropriate regression model to  employ, a clear technique 
should be used such as the Akaike information criterion or 
Bayesian information criterion, and ideally selected previous to seeing the results.
Also, to ensure a reasonable fit, the scientist should also make regression quality plots, 
such as residual plots,  and not just rely on a single criterion 
such as reduced $\chi^2$.
When we  apply these techniques to low four-momentum transfer cross section data, we find a proton radius 
that is consistent with the muonic Lamb shift results.  While presented for the case of proton radius extraction,
these concepts are applicable in general and can be used to  illustrate the necessity of balancing bias and variance 
when building a regression model and validating results, ideas that are at the heart of modern machine learning algorithms.
\end{abstract}

\maketitle

\section{Introduction}

High-precision Lamb shift experiments on muonic hydrogen atoms have determined the proton radius to 
be 0.84087(39)~fm~\cite{Pohl:2010zza,Antognini:1900ns}.   This result is in stark contrast to the
CODATA-2014 recommended value of 0.8751(61)~fm~\cite{Mohr:2015ccw} which comes 
from electron scattering results and atomic transition frequencies~\cite{Boshier:1989zz,Weitz:1994zz,Berkeland:1995dyd,Bourzeix:1996zz,Udem:1997zz}.
The discrepancy between these radius values has become known as the proton 
radius puzzle~\cite{Pohl:2013yb,Carlson:2015jba,Gao:2015aax,Pohl:2016tqq,Nez:2011zz,Krauth:2017ijq}.
While initial efforts to understand this puzzle focused on the details of the muonic experiment, attention has
now turned to re-examining the atomic and electron scattering 
results~\cite{Kelkar:2016tcx,Beyer79,fleurbaey:tel-01633631,Fleurbaey:2018fih}.   

For the electron scattering data, the proton charge radius, $r_p$, is extracted from
cross section data by determining the slope of the electric form factor, $G_E$, in the
limit of four-moment transfer, $Q^2$, approaching zero: 
\begin{equation}
\label{eq:radius}
  r_p \equiv 
    \left( -6  \left. \frac{dG_E(Q^2)}{dQ^2}
    \right|_{Q^{2}=0} \right)^{1/2} \>.
\end{equation}
This definition of the radius has been shown to be the same as the one used by the Lamb 
shift measurements~\cite{Miller:2018ybm}.
Unfortunately, electron scattering experiments cannot reach the $Q^2 = 0$ limit; thus,
an extrapolation is required to determine the charge radius from the experimental data.

\begin{figure}[htb]
\includegraphics[width=\columnwidth]{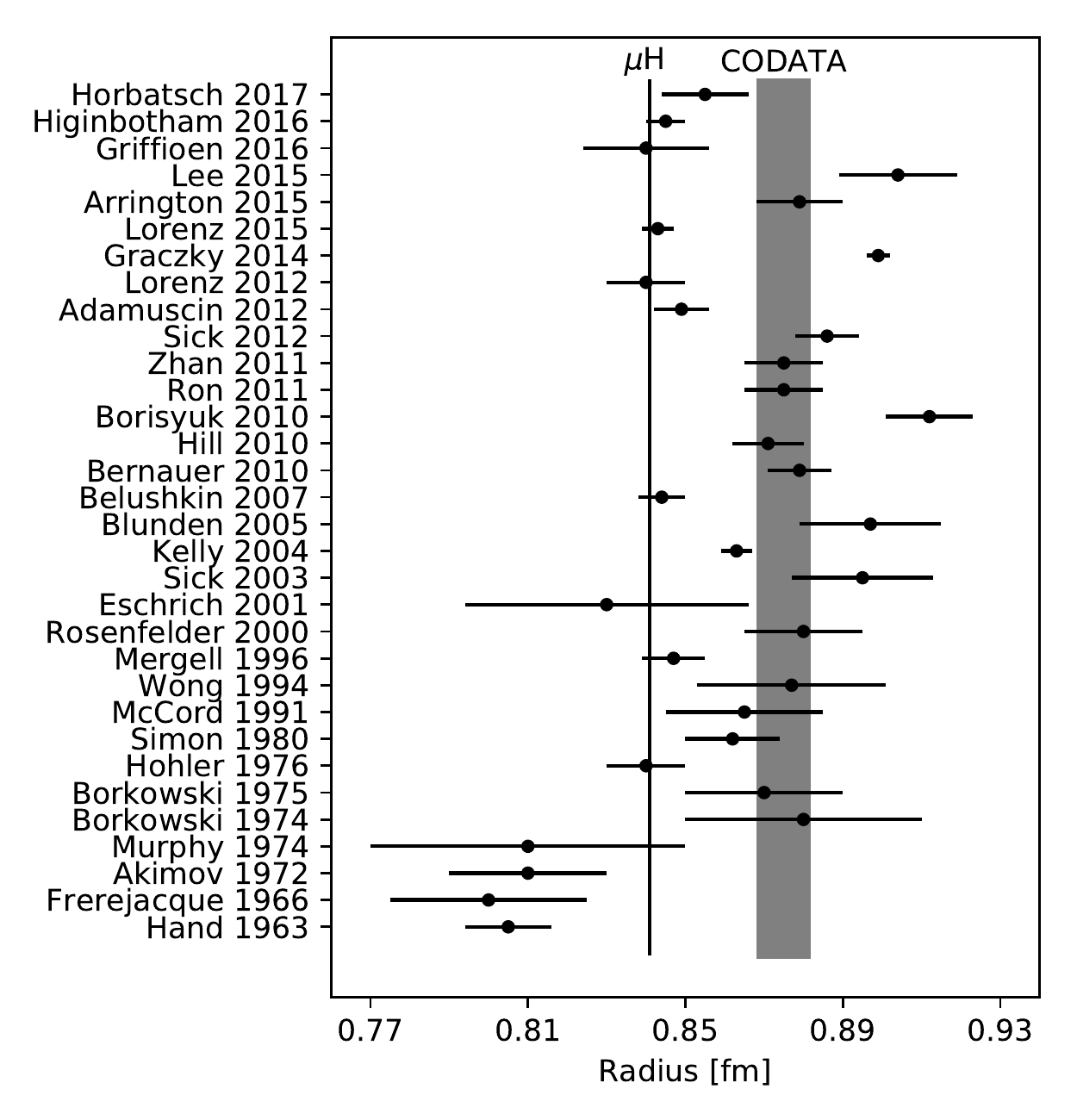}
\caption{Shown are the radii extracted from electron scattering data~\cite{Horbatsch:2016ilr,
Higinbotham:2015rja, Griffioen:2015hta, Lee:2015jqa, Arrington:2015ria, Graczyk:2014lba,
Lorenz:2014yda,Lorenz:2014vha, Lorenz:2012tm, Adamuscin:2012zz, Sick:2012zz, Zhan:2011ji,
Ron:2011rd, Borisyuk:2009mg, Hill:2010yb, Bernauer:2010wm, Bernauer:2013tpr, Belushkin:2006qa,
Blunden:2005jv, Kelly:2004hm, Sick:2003gm, GoughEschrich:2001ji, Rosenfelder:1999cd, Mergell:1995bf,
Wong:1994sy, McCord:1991sd, Simon:1980hu, Hohler:1976ax, Borkowski:1975ume, Borkowski:1974tm,Borkowski:1974mb,
Akimov:1972nu, Murphy:1974zz, Frerejacque:1965ic, Hand:1963zz}.
The vertical bands indicate the value and uncertainty of the
proton radius from muonic hydrogen~\cite{Pohl:2010zza,Antognini:1900ns} and 
CODATA-2014~\cite{Mohr:2015ccw}.
}
\label{collection}
\end{figure}

The various proton radius values that have been extracted from electron scattering data
are shown in Fig.~\ref{collection}.    
In general, $r_p$ extractions using simple statistical modeling of the 
low $Q^2$ data~\cite{Horbatsch:2015qda,Rosenfelder:1999cd,Griffioen:2015hta,Higinbotham:2015rja}
tend towards a smaller radius while the more complex statistical model, with many free 
parameters, tend to extract a large proton 
radius~\cite{Bernauer:2010wm,Bernauer:2013tpr,Lee:2015jqa,Graczyk:2014lba,Lorenz:2014vha}.
Nuclear theory constrained extractions of the radius tend to 
favor a smaller radius~\cite{Hohler:1976ax,Belushkin:2006qa,Horbatsch:2016ilr}.
One technical detail that is affecting all 
these results is how the normalization of the experimental data is handled.

In general, regardless of the extracted radius, these different fits
tend to agree rather well at moderate $Q^2$ and the differences between functions
can only be clearly seen examining the very low $Q^2$ region where slope
and normalization are systematically linked together.
A general criticism of the small radius extractions of the proton radius is 
the presence of statistical bias~\cite{Sick:2017aor,Sick:2018fzn},
with an implication that bias needs to be avoided in order to 
successfully extract the true radius from the data.   
The use of Monte Carlo methods to find bias and then reject simple proton radius extraction methods
originated with the classic Monte Carlo study of Borkowski~{\it{et al.}}~\cite{Borkowski:1975ume} 
where linear extrapolations were flatly rejected in favor of quadratic extrapolations. 
Interestingly, that work ignored the variance of the more complex function.

We will show in this work that when using a Monte Carlo study to test a model's ability
to extract the proton radius one needs to consider not only bias but also variance,
and find an appropriate balance between the two. After all, it is better to have a slightly biased watch to tell the hour, than a broken one that is unbiased because it overestimates and underestimates the time symmetrically.

We will also illustrate that one must consider 
the range, quantity and precision of the data
when determining the best predictive statistical model
and show that simple biased statistical models can have a higher predictive validity than unbiased more complex models~\cite{Shmueli:2010}.   We will then 
apply these ideas to model selection with real
data, where instead of millions of Monte Carlo results, 
we get but a single realization of the possible outcomes.  

\section{Bias}

In the English language, bias is often used as a pejorative term. 
In the context of regression, it is simply an offset of the mean from the true central value.
Since it is part of a distribution, it
is not a property of a single realization but can be determined by repeated sampling.    In the context of the proton 
radius extractions, bias was nicely illustrated by Borkowski~{\it{et al.}}~\cite{Borkowski:1975ume} and we will
describe their procedure in the following paragraphs.

Form factor pseudo-data is first systematically generated 
from $\unit[0.1]{fm^{-2}}$ to 0.4, 0.8, 1.2,
and $\unit[1.6]{fm^{-2}}$ 
in steps of $\unit[0.05]{fm^{-2}}$ 
using the standard dipole function:
\begin{equation}
\label{sd}
G\mathrm{_D}(Q^2) = ( 1 + Q^2/(\unit[18.23]{fm^{-2}}))^{-2},
\end{equation}
where the cutoff parameter of 18.23~fm$^{-2}$ corresponds to a radius of 0.8113~fm.

Next, to mimic real data, the pseudo data points are each randomly shifted using a 
normal distribution with mean zero and a sigma of 0.5\%. 
Then the pseudo data sets are then fit using both linear and quadratic functions:
\begin{align}
f_{\mathrm{linear}}(Q^2) &  = a_0 + a_1 Q^2, \label{Eq:linear} \\
f_{\mathrm{quadratic}}(Q^2) & = a_0 + a_1 Q^2 + a_2 Q^4. \label{Eq:quadratic}
\end{align}
These functions are written so that they are linear in the fit coefficients,
which
allows the $\chi^2$ minimization to be performed exactly
while also allowing the normalization to float.
To obtain the physical G$_E$(0) = 1 behavior, one simply divides these functions 
by the normalization term, $a_0$, to find the slope of G$_E$(0) given by $a_1/a_0$
which can be used in Eq.~\ref{eq:radius} to determine the radius.

This procedure was repeated with $10^6$ sets of pseudo data to 
precisely determine the mean of the extracted 
radii for these two functions.   Since the standard dipole was used as the input function, one would expect an unbiased 
function to return a radius of 0.8113~fm.
Table~\ref{ztable} reproduces the original result~\cite{Borkowski:1975ume}.
As the table shows, the mean values of $a_0$ and $r_p$ show a clear bias. 
Based on this study, the authors of the original work erroneously concluded 
that the linear models should always be rejected in favor of the lower-bias 
quadratic function.

\begin{table}
\caption{The mean $a_0$ and radius from doing $10^6$ Monte Carlo simulations
for each interval in $Q^2$
where Eq.~\ref{sd} was used to generate pseudo data in $\unit[0.05]{fm^{-2}}$ steps
with each data point smeared by a randomly generated, normally distributed point-to-point 
uncertainty of 0.5\%.
The results clearly indicate that the linear fits are biased.   The input
radius was \unit[0.8113]{fm} (an $a_1/a_0$ term of $\unit[0.1097]{fm^{-1}}$) and $a_0 = 1$.}
\begin{tabular}{c|cc|cc} \hline
interval       & \multicolumn{2}{c|}{linear fit} & \multicolumn{2}{c}{quadratic fit}  \\
fm$^{-2}$      & $a_0$      & Radius [fm]          & $a_0$    & Radius [fm] \\ \hline
 0.1 -- 0.4 & 1.000& 0.79& 1.000& 0.81 \\
 0.1 -- 0.8 & 0.999& 0.78& 1.000& 0.81 \\
 0.1 -- 1.2 & 0.997& 0.77& 1.000& 0.81 \\
 0.1 -- 1.6 & 0.996& 0.76& 1.000& 0.81 \\ \hline
\end{tabular}
\label{ztable}
\end{table}

\section{Variance}

\begin{figure}[htbp]
\includegraphics[width=\columnwidth]{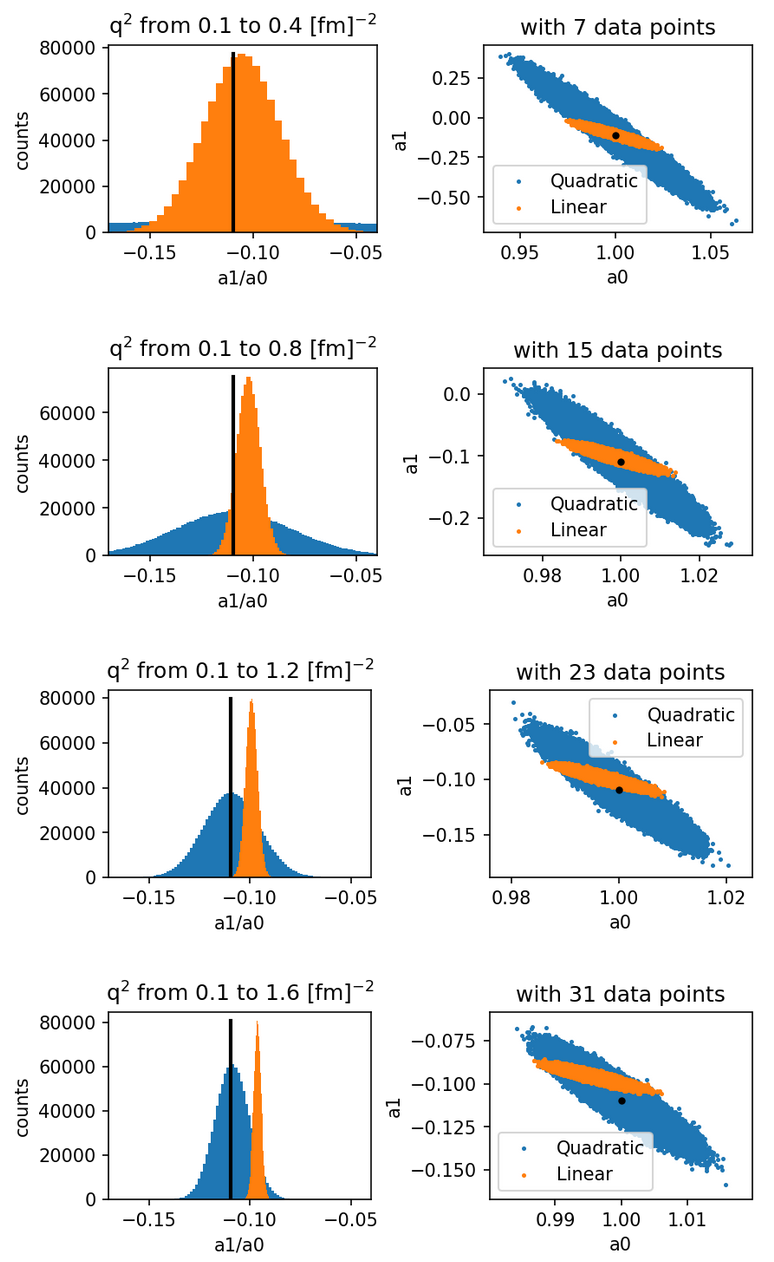}
\caption{A graphic representation of the Monte Carlo results showing how the linear fits tend to have a relatively
high bias though a low variance, while the quadratic fits tend to have a relatively low bias but a large variance. The black line/point represents the true value.}
\label{mc-results}
\end{figure}

While the linear fit does exhibit a bias, bias is not the only quantity that must
be considered when selecting an appropriate statistical model to use.
In particular, along with the offset from the true value, bias, one must also consider the 
distribution of the outcomes, the variance.   Where variance is the square of
the standard deviation, $\sigma$, of the statistical distribution.
Table~\ref{fulltable} shows a more complete picture of the simulation results 
where the $\sigma$ of the results is shown along with the bias and is
graphically represented in Fig.~\ref{mc-results}. 

\begin{table*}
\caption{An expanded version of Table~\ref{ztable} where instead of just showing the mean offset of the 
fit results for $a_1/a_0$, the bias, we also indicate the width of the fit results, $\sigma$. Recall that the point uncertainty is fixed at 0.5\%.  For the simulated radius of 0.8113~fm,
one would expect an unbias fit to give $a_1/a_0$ of 0.1097~fm$^2$; thus, the difference of that value from the  mean fitted 
value of $a_1/a_0$ is the bias and the width of the distribution, $\sigma$.   Also shown is the
root mean square error, RMSE, which can be used to quantify the best function for a given interval taking into account both
bias and variance.}
\begin{tabular}{cc|cccccc|cccccc} \hline
       &           & \multicolumn{6}{c|}{linear fit}                       & \multicolumn{6}{c}{quadratic fit}                    \\ 
Data   & Interval  & $a_0$ & Radius &  $a_1/a_0$ &  Bias  & $\sigma$ &  RMSE  & $a_0$ & Radius & $a_1/a_0$ &  Bias & $\sigma$ &  RMSE \\  
Points & fm$^{-2}$ &       & [fm]   &   [fm$^2$] &[fm$^2$]&[fm$^2$]&[fm$^2$]&       & [fm]   &  [fm$^2$]&[fm$^2$]&[fm$^2$]&[fm$^2$]\\  \hline
7      & 0.1 -- 0.4 & 0.9995& 0.7949& $-0.1053$& $-0.0044$& 0.0184& 0.0189 & 1.0000& 0.8065& $-0.1084$& $-0.0013$& 0.1094& 0.1094\\
15     & 0.1 -- 0.8 & 0.9987& 0.7827& $-0.1021$& $-0.0076$& 0.0057& 0.0095 & 1.0000& 0.8094& $-0.1092$& $-0.0005$& 0.0281& 0.0281\\
22     & 0.1 -- 1.2 & 0.9975& 0.7711& $-0.0991$& $-0.0106$& 0.0030& 0.0110 & 0.9999& 0.8087& $-0.1090$& $-0.0007$& 0.0138& 0.0138\\
31     & 0.1 -- 1.6 & 0.9959& 0.7601& $-0.0963$& $-0.0134$& 0.0019& 0.0136 & 0.9998& 0.8075& $-0.1087$& $-0.0010$& 0.0085& 0.0085\\ \hline
\end{tabular}
\label{fulltable}
\end{table*}

\begin{table*}
\caption{Same as Table~\ref{fulltable}, but now with equal number of data points for each range.}
\begin{tabular}{cc|cccccc|cccccc} \hline
       &           & \multicolumn{6}{c|}{linear fit}                       & \multicolumn{6}{c}{quadratic fit}                    \\ 
Data   & Interval  & $a_0$ & Radius &  $a_1/a_0$ &  Bias  & $\sigma$ &  RMSE  & $a_0$ & Radius & $a_1/a_0$ &  Bias   & $\sigma$ &  RMSE   \\  
Points & fm$^{-2}$ &       & [fm]   &[fm$^2$]    &[fm$^2$]&[fm$^2$]&[fm$^2$]&      &[fm]    &[fm$^2$]   & [fm$^2$]&[fm$^2$]& [fm$^2$] \\  \hline
31& 0.1 -- 0.4 & 0.9995& 0.7952& $-0.1054$& $-0.0043$& 0.0098& 0.0107 & 1.0000& 0.8091& $-0.1091$& $-0.0006$& 0.0629& 0.0629 \\
31& 0.1 -- 0.8 & 0.9987& 0.7827& $-0.1021$& $-0.0076$& 0.0041& 0.0086 & 1.0000& 0.8098& $-0.1093$& $-0.0004$& 0.0208& 0.0208  \\
31& 0.1 -- 1.2 & 0.9974& 0.7711& $-0.0991$& $-0.0106$& 0.0026& 0.0109 & 0.9999& 0.8091& $-0.1091$& $-0.0006$& 0.0121& 0.0121  \\
31& 0.1 -- 1.6 & 0.9959& 0.7601& $-0.0963$& $-0.0134$& 0.0019& 0.0136 & 0.9998& 0.8075& $-0.1087$& $-0.0010$& 0.0085& 0.0085  \\  \hline
\end{tabular}
\label{equaldatatable}
\end{table*}

For all $Q^2$ intervals, the linear fits provide significantly smaller $\sigma$ than the more complex quadratic fits. Though picking the function to use
based solely on variance would also be incorrect.  
Thus, this seemingly simple example has turned into a nearly textbook illustration 
of the trade-off between variance and bias with the linear fits having a relatively 
high bias with a low variance, while the quadratic fits have a low bias and high variance.
Of course, to calculate the bias requires we know the true value which often isn't the case
in a real experiment.
%
%

\section{Goldilocks Dilemma}\label{Goldilocks}

\begin{figure}
\includegraphics[width=\columnwidth]{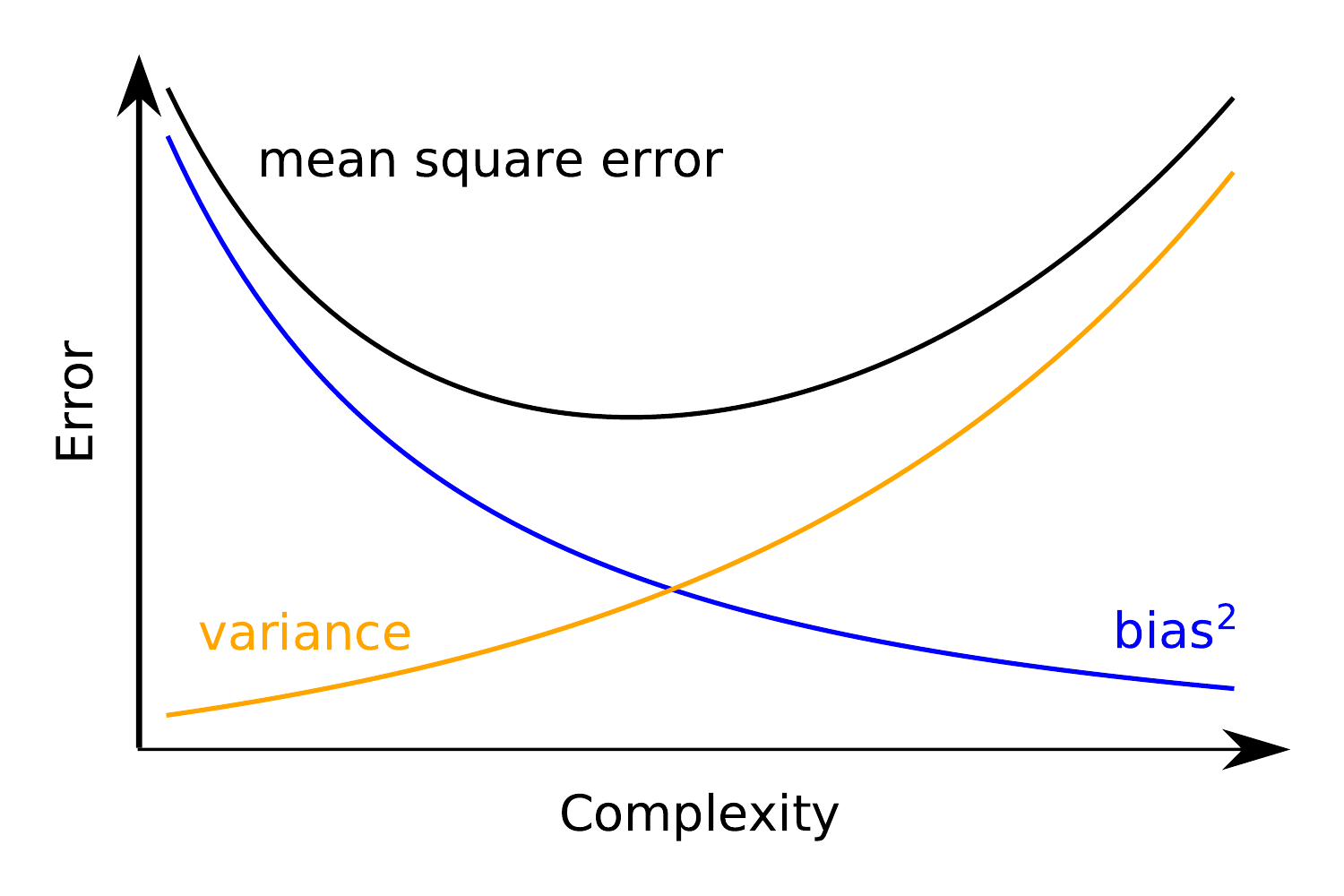}
\caption{An illustration of the trade-off between bias and variance when selecting a statistical model.   Simple models
will have low variance but high bias (under-fitting) while complex models will have low bias but high variance (over-fitting).   
It is this trade-off that one seeks to balance.   While with repeated  Monte Carlo simulations it is trivial to find the optimal
predictive model for a given set of data, in the real world the true model is typically unknown; one only gets to perform
a very limited number of experiments and thus, one relies on using real data and statistical methods for 
model selection~\cite{Hastie:2009}.}
\label{biasvariance}
\end{figure}

As was noted by George Box, all models are wrong, thus, the goal is to find the most useful model~\cite{Box76}. For any given set of statistical models, the goal is to find the optimal balance between bias and variance.
In general, this can be written as
\begin{equation}
\frac{d {\mathrm{Bias}^2 }}{ d {\mathrm{Complexity}}} \approx \frac{- d {\mathrm{Variance}} }{ d {\mathrm{Complexity}}},
\end{equation}
as illustrated in Fig.~\ref{biasvariance}.
Thus, to quantify the goodness of the fits, we choose Root Mean Square Error (RMSE)~\cite{Hastie:2009,James:2014},
\begin{equation}\label{RMSEdef}
{\mathrm{RMSE}} = \sqrt{ {\mathrm{bias}}^2 + {\mathrm{Sigma}}^2} = \sqrt{{\mathrm{Bias}}^2 + {\mathrm{Variance}}}.
\end{equation}

\begin{figure}[htbp]
\includegraphics[width=\columnwidth]{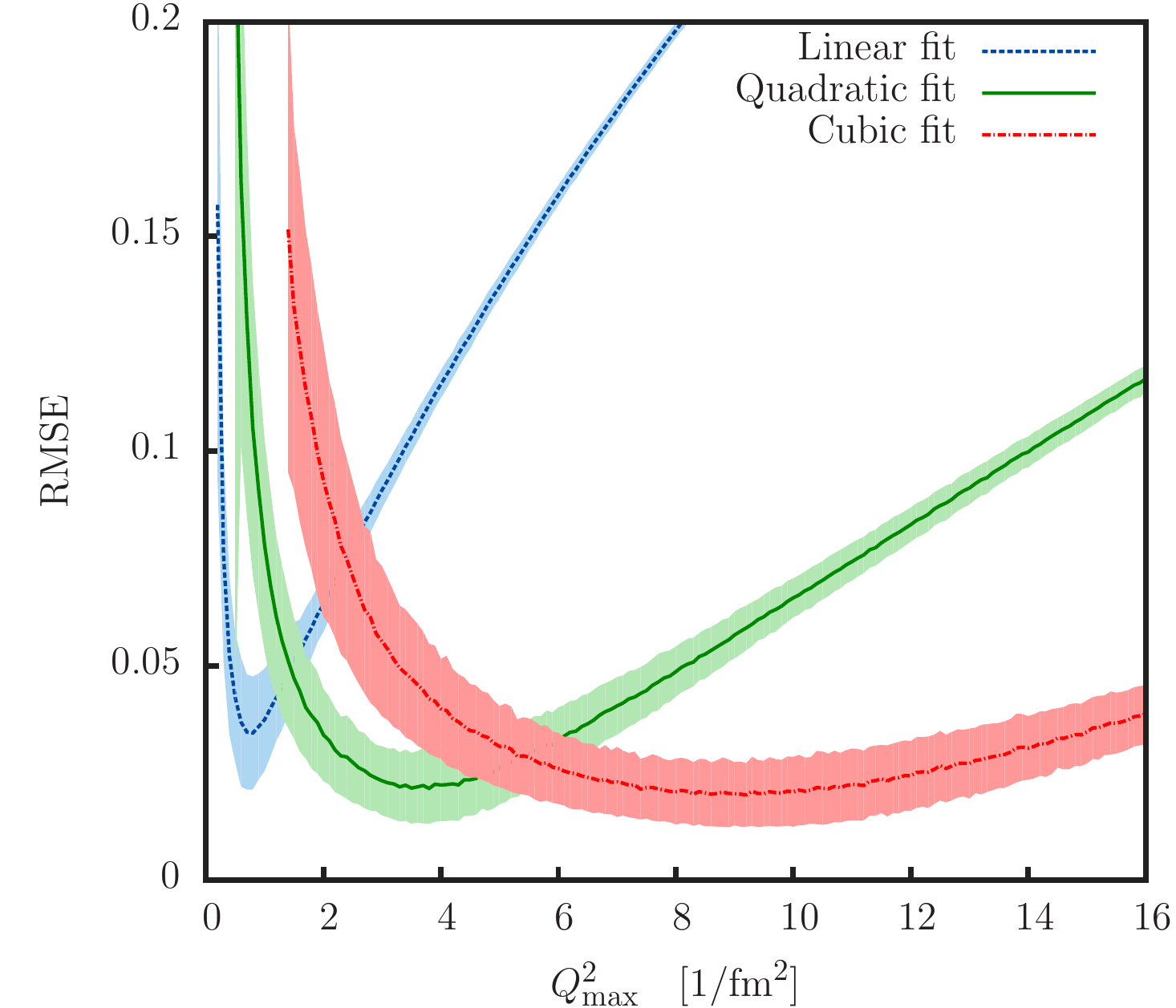}
\caption{A graphic showing how as one goes to higher and higher $Q^2$ 
with the same point-to-point uncertainty and spacing,
        an ever more complex model is needed to represent the data.}
\label{mihaplot}
\end{figure}

Using the RMSE values in Table~\ref{fulltable}, one can now quantify 
that for this example the 0.1--0.8~fm$^{-2}$ interval is the preferred range for the linear model 
while the 0.1--1.6~fm$^{-2}$ interval is the preferred range for the quadratic model.   Going to even
higher $Q^2$ will require even more complex models as illustrated in Fig.~\ref{mihaplot}.
This is in contrast to the conclusion one draws when one only considers bias as presented in Table~\ref{ztable},
though consistent with the observation that the optimal specific form of the parameterization
may depend on the $Q^2$ region being fit~\cite{Alberico:2008sz}.

\begin{figure}
\includegraphics[width=\columnwidth]{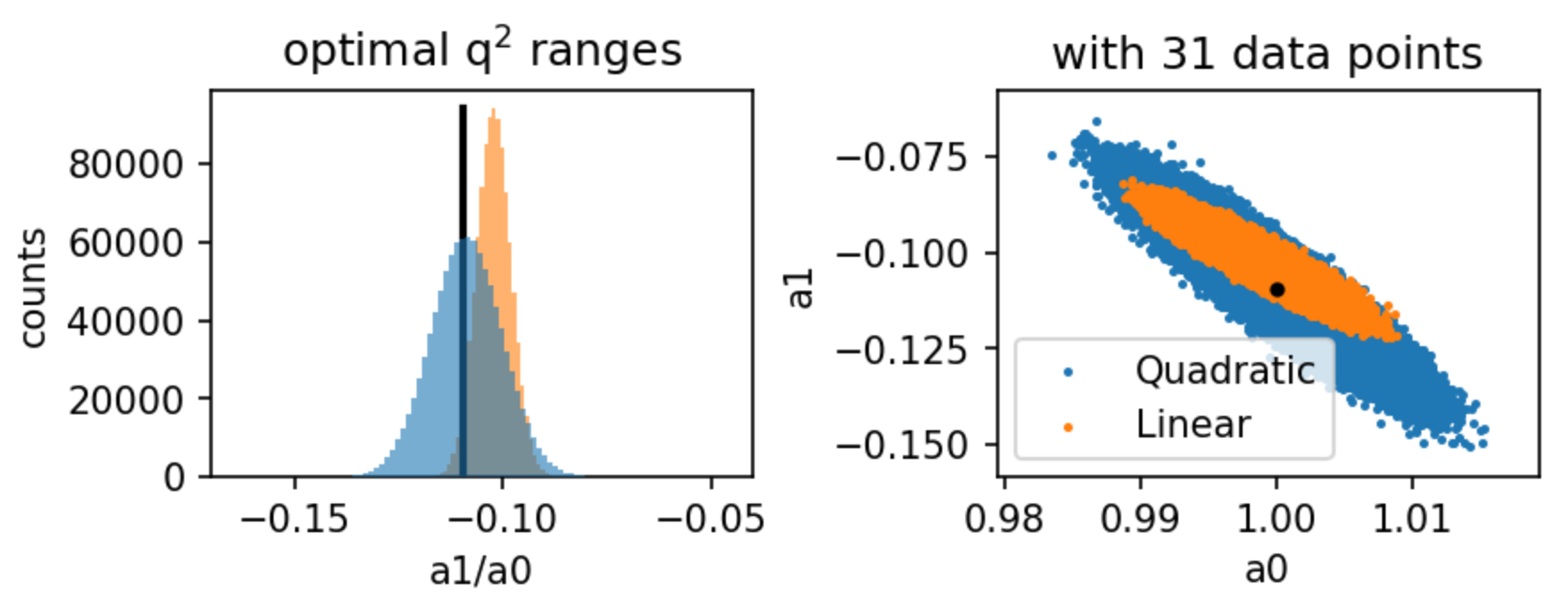}
\caption{The result of a million simulations and fits of linear fits over the $Q^2$ range 
of 0.1 -- 0.8~fm$^{-2}$ and quadratic fits over 0.1 -- 1.6~fm$^{-2}$, both with 31 uniformly 
spaced data points.    Using root mean
square error as the metric, neither example is significantly better than the other for exacting the proton
radius.   This is analogous to a dart game between two equally skilled players: one who hits the bull's eye more 
often yet has a large spread (low bias but high variance), and another, equally skilled, player who has a 
tighter cluster of hits but an offset (high bias but low variance).}
\label{zoptimized}
\end{figure}

It is interesting to repeat the Monte Carlo simulation for equal number of data points within each range
especially since, for elastic scattering, cross sections are significantly higher at lower values of $Q^2$
and thus, it is easy to obtain more low $Q^2$ data.
This is shown in Table~\ref{equaldatatable} and now the picture is even grayer as the RMSE of the linear 
fit is nearly equal to the quadratic, thus, assuming that the standard dipole was the true generating function, an experiment
with 31 data points and an uncertainty of 0.005 per point over a range of 0.1 to 0.8~fm$^{-2}$ and a different experiment over a range of 0.1 to 1.6~fm$^{-2}$ would have an equal probability of reproducing the correct radii if all other things were equal (given that the modeler in the first group adjusts a line and the one in the second a parabola). 
This is visualized in Fig.~\ref{zoptimized} where the linear fit is clearly biased but has a small variance compared to
the unbiased, large variance quadratic fit.

The choice of the parsimonious modeler to use the low $Q^2$ data would likely be 
driven by the recognition of the fact that as $Q^2$ increases
the extraction of the charge form factor is complicated by the growing influence of the magnetic 
form factor.   The choice to use a larger $Q^2$ range would likely be driven by a desire 
to form a more complete picture of the proton's structure.
For example, the parsimonious modeler may only be interested in the proton radius while another 
modeler may be interested in higher order moments~\cite{Alarcon:2017lhg}.
Thus, the tension in the extractions of the proton radius from electron scattering data is really 
about the fact that modelers using the low $Q^2$ are generally getting a systematically different 
result than the modelers doing fits which include high $Q^2$ data, and perhaps points to a systematic 
problem with our knowledge of the magnetic form factor and/or the functional form of the
form factors.

\section{The Best Predictive Model}

Selecting between a linear or quadratic regression of the more complex standard dipole function may seem 
a bit contrived, as one might naively think that just using the generating function itself would always yield the
best results. In section~\ref{SemiAnalytical} we prove that for non linear fits there is an induced bias even when using the generating function.

When taking into account the variance, we can also show that this is not always the case with a simple set up: 
we use the lowest $Q^2$ range, 0.1 -- 0.4 fm$^{-2}$, and replace the quadratic function 
with the generating function and a floating normalization term:
\begin{equation}
\label{eq:fitdipole}
f_{{\mathrm{Dipole Fit}}}(Q^2) =  n_0 ( 1 - b_1 Q^2 / 2)^{-2},
\end{equation}
where $n_0$ is the normalization factor and the radius is given by $\sqrt{-6 b_1}$.
Pseudo data is generated for absolute random errors of 0.01, 0.005 and 0.003 
with three different spacings: 
0.05 fm$^{-2}$ spacing with 7 points, 0.01 fm$^{-2}$ spacing with 31 points, and 
0.005 fm$^{-2}$ spacing with 61 points.       
The results of fitting these pseudo data sets are shown in Table~\ref{simpleVSperfect}.    
For a given row in the Table, the linear fit has the greater bias and the dipole fit always
has the greater variance; bringing the root mean square error very close for all the test cases.   
This example also makes it clear that it is not just the number of points that matter, 
but the size of the uncertainties and the range of the data that will also be critical parameters
in model selection. In the cases where the MSE of the line was smaller than the ones of the Dipole, the line was being a better ``predictive'' model for the proton radius, while the Dipole was a better ``descriptive'' model of the form factor as whole.

Of course for real data, nature hides the true generating function from us, so perhaps it is reassuring to know
that a reasonable approximation is able to reveal the underlying physics just as well as, if not better than, the
true function.   
To be clear, the lesson is not that one function is better than another; it is that for a given set of data,
the scientist is challenged to use the most appropriate model (either descriptive or predictive) for
the task at hand.   Further details on the general mathematics behind these example problems can be 
found in~\cite{Shmueli:2010}.    

\begin{table*}
\caption{For the lowest Q$^2$ interval, 0.1 to 0.4 fm$^{-2}$, we compare regressions with
a linear function (Eq.~\ref{Eq:linear}) to the dipole fit function (Eq.~\ref{eq:fitdipole}).   Keeping the range
fixed, a spacing of 0.05~fm$^{-2}$ (7 points), 0.01~fm$^{-2}$ (31 points) and 0.005~fm$^{-2}$ (61 points)
was used with various absolute random errors.  In several of these cases, the simple linear 
function provides a better predictive model then the true functional form and is never 
far from the true function.}
\begin{tabular}{cc|cccccc|cccccc} \hline
       &          & \multicolumn{6}{c|}{linear fit}                       & \multicolumn{6}{c}{dipole fit}                     \\ 
Data   & Random   & $a_0$ & Radius &  $a_1/a_0$ &  Bias  & $\sigma$ &  RMSE  & $a_0$ & Radius & $a_1/a_0$ &  Bias & $\sigma$ &  RMSE \\  
Points & Error    &       & [fm]   &   [fm$^2$]& [fm$^2$]&[fm$^2$]&[fm$^2$] &       & [fm] &[fm$^2$]& [fm$^2$] &[fm$^2$]&[fm$^2$] \\  \hline
7      & 0.01     & 0.9995& 0.7941& $-0.1051$& $-0.0046$& 0.0359& 0.0361 & 1.0001& 0.8109& $-0.1096$& $-0.0001$& 0.0378& 0.0378  \\ 
7      & 0.005    & 0.9995& 0.7945& $-0.1052$& $-0.0045$& 0.0174& 0.0180 & 1.0000& 0.8113& $-0.1097$& $0.0000$& 0.0194& 0.0194  \\
7      & 0.003    & 0.9996& 0.7952& $-0.1054$& $-0.0043$& 0.0108& 0.0116 & 1.0000& 0.8113& $-0.1097$& $0.0000$& 0.0114& 0.0114  \\ \hline
31     & 0.01     & 0.9996& 0.7945& $-0.1052$& $-0.0045$& 0.0186& 0.0191 & 1.0000& 0.8113& $-0.1097$& $0.0000$& 0.0207& 0.0207  \\
31     & 0.005    & 0.9995& 0.7952& $-0.1054$& $-0.0043$& 0.0093& 0.0102 & 1.0000& 0.8113& $-0.1097$&  0.0000& 0.0103& 0.0103  \\
31     & 0.003    & 0.9995& 0.7952& $-0.1054$& $-0.0043$& 0.0056& 0.0070 & 1.0000& 0.8113& $-0.1097$&  0.0000& 0.0062& 0.0062   \\ \hline 
61     & 0.01     & 0.9995& 0.7949& $-0.1053$& $-0.0044$& 0.0135& 0.0142 & 1.0000& 0.8113& $-0.1097$&  0.0000& 0.0150& 0.0150  \\ 
61     & 0.005    & 0.9995& 0.7952& $-0.1054$& $-0.0043$& 0.0069& 0.0081 & 1.0000& 0.8113& $-0.1097$&  0.0000& 0.0073& 0.0073  \\ 
61     & 0.003    & 0.9995& 0.7952& $-0.1054$& $-0.0043$& 0.0040& 0.0059 & 1.0000& 0.8113& $-0.1097$&  0.0000& 0.0045& 0.0045  \\ \hline 
\end{tabular}
\label{simpleVSperfect}
\end{table*}

\begin{figure}[htb]
\includegraphics[width=\columnwidth]{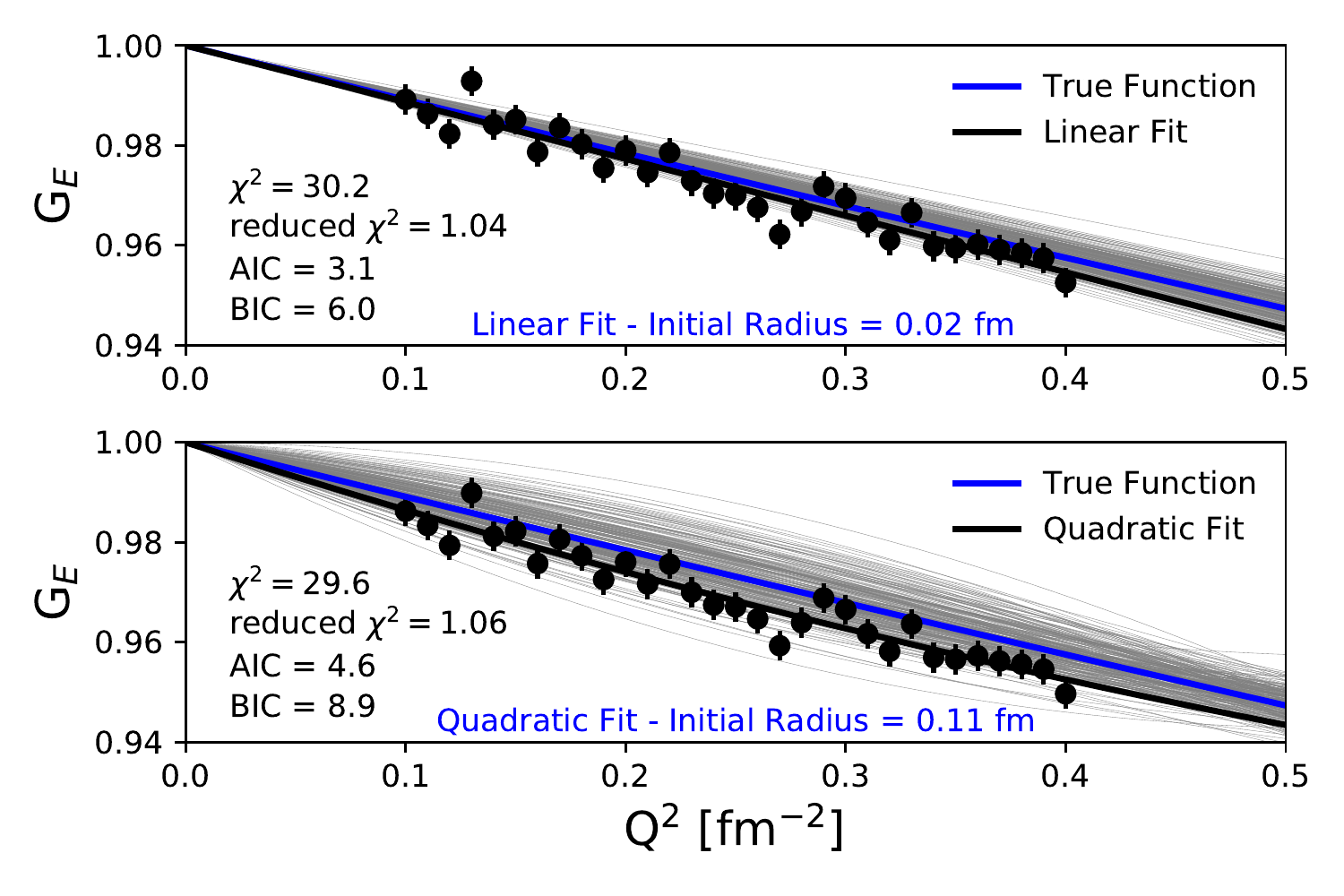}
\caption{Illustration of the effect of renormalizing data.  The
black points show one set of pseudo data, with absolute 0.3\%  random point-to-point uncertainty and 0.01 fm$^{-2}$ spacing that have been renormalized
by applying the prior that $G_E(0)=1$.    While the prior is true, using an inappropriate model
can cause both the extracted radius and normalization to be dramatically shifted from the true
function.   
The grey bands were created by preforming of 250 simulations and are presented as
an animation in the supplemental material~\cite{Supplement}.}
\label{linearVSquadratic}
\end{figure}

\begin{figure}[htb]
\includegraphics[width=\columnwidth]{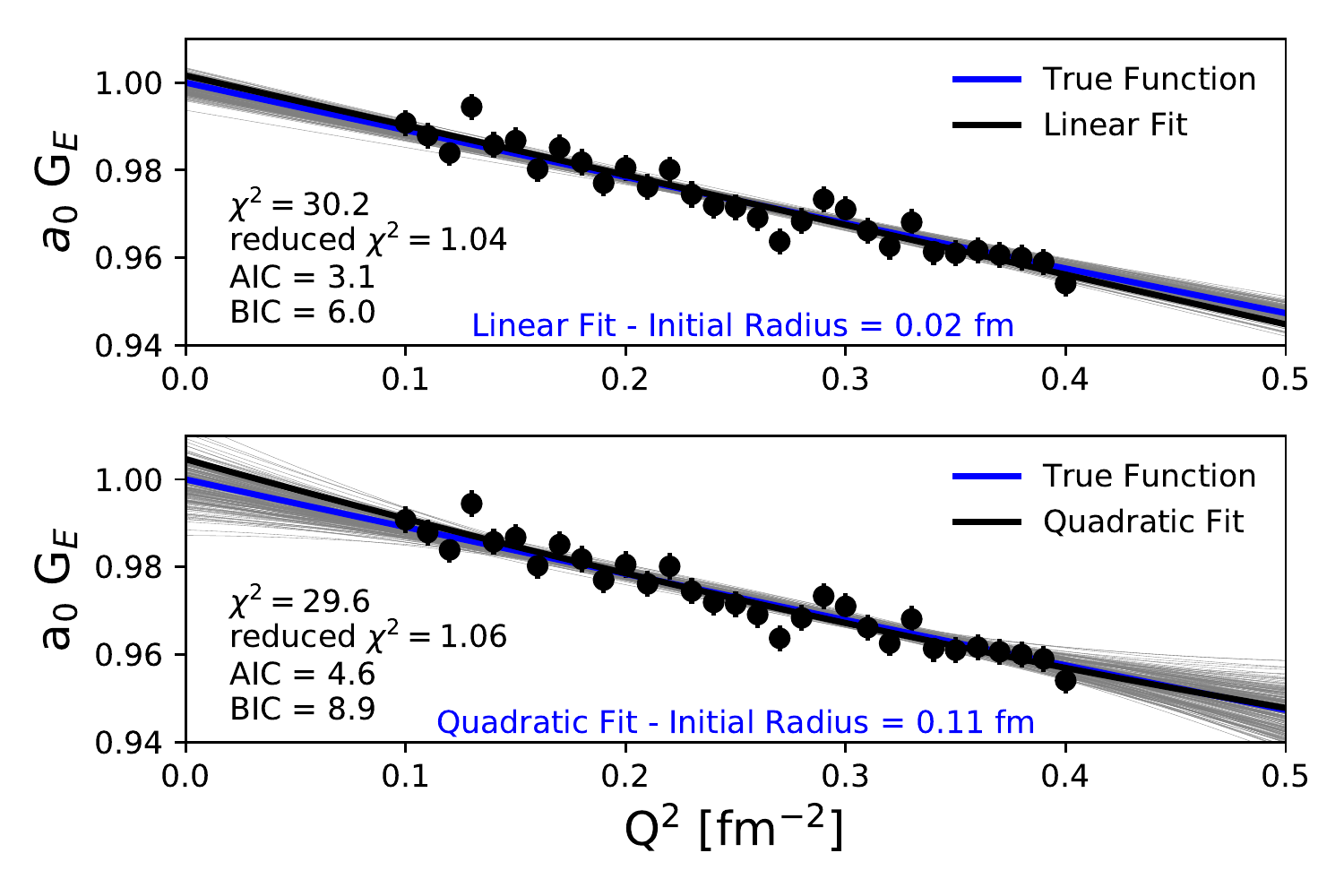}
\caption{Instead of immediately applying the prior that $G_E(0)=1$ as done in Fig.~\ref{linearVSquadratic}, 
it is more intuitive to simple fit the data allowing the end-point to float.  
In fact, a frequentist would correctly simply check that the function reasonably points to one within
the normaliztion uncertainty.
The grey bands were created using the same pseudo data 
as used in Fig.~\ref{linearVSquadratic}. 
and are presented as
an animation in the supplemental material~\cite{Supplement}. 
While these two figures (Fig. 7 and 8) may look very different, as far as the extracted radius goes, 
these two different ways of presenting the pseudo data
give exactly the same results.}
\end{figure}

Using an inappropriate model can lead to erroneous conclusions (e.g. due to different normalziations of the data). 
In Fig.~\ref{linearVSquadratic} we show an
example of one set of the pseudo data fit with a linear and quadratic where the prior that the
$G_E(0)=1$ has been applied (i.e. the data has been divided by the normalization term from the regression
so the function goes to the known limiting value at the origin).
Since here we know the function that generated the pseudo data, it is clear that the quadratic
fit can cause the data to be inappropriately shifted and can generate an large variations
in the radius.
Of course in the real world, the true function is not known; thus, with real data
a model selection technique is required in order to determine the appropriate function. 

\section{Model Selection}

While this classic Monte Carlo example problem is over 40 years old, it actually points to exactly the split in 
the current electron scattering proton radius extraction procedures.     The parsimonious modelers, who are 
focused solely on extracting a radius, have focused on the low $Q^2$ region accepting a slightly higher 
level of bias in exchange for low variance; on the other hand, those modelers who are interested in extracting more information 
about the proton (e.g. higher order moments) fit longer $Q^2$ ranges and have
focused on complex models which, while lower in bias, come at a cost of higher variance.  

Also, since we do not know the true model, one cannot in general calculate the RMSE. So, while Monte Carlo exercises 
like the one described herein are extremely useful for finding reasonable models to consider and understanding
expected uncertainties, the data must be used to select the appropriate model. 
For this one can rely on statistical modeling selection techniques such as an $F$-test for
nested models~\cite{Bevington:2003,James:2006,Sirca:2016} or the more general Akaike information criterion (AIC)~\cite{Akaike:1974} 
or Bayesian information criterion (BIC)~\cite{Schwarz:1978} 
to guide our selection of the most appropriate model to describe a given set of data.
These statistical criteria are calculated as follows: 
\begin{align}
\chi^2         & = \sum_{n=1}^{N}((\mathrm{data_i} - \mathrm{model}) / \sigma_i)^2, \\
reduced~\chi^2 & = \chi^2/ (N - N_{\mathrm{var}}),  \\
\mathrm{AIC}            & = N \log(\chi^2/N) + 2 N_{\mathrm{var}}, \\
\mathrm{BIC}            & = N \log(\chi^2/N) + \log(N) N_{\mathrm{var}},
\end{align}
where $N$ is the number of data points, $data_i$ and $\sigma_i$ are measured values and estimated uncertainties,
and $N_{\mathrm{var}}$ is the number of model parameters.
Further details about current model selection techniques can be found in~\cite{Ernst:2012}.

One should also keep in mind that the input models in Monte Carlo simulations are always just an approximation
and one needs to be careful about drawing too strong inferences from the simulated results. 
For example, just because the linear model has a negative bias when compared to the standard dipole, 
does not imply that it has a negative bias with respect to all possible models.

\section{Real Data}

This brings us to the real data and the current proton radius extractions.   Many different functions have
been tried over the years from simple linear fits~\cite{Hand:1963zz,Murphy:1974zz} and continued fractions~\cite{Sick:2003gm} 
to high order polynomials with~\cite{Bernauer:2013tpr} and without constraints~\cite{Lee:2015jqa}.
Since obtaining sub-percent
level absolute cross sections is nearly impossible, a normalization parameter is included to allow an entire set of data to shift
as was done in the Monte Carlo simulations.   Again, this just allows that the prior, $G_E(0)=1$, can be applied.
To be clear, in this work, when referring to a set of experimental data we are referring to a group of data
with a single normalization parameter.

The measured cross sections, $\sigma_{\mathrm{Meas}}$,  are related to the charge and electric form factors via 
\begin{equation}
\frac{\sigma_{\text{Meas}}}{\sigma_{\text{Mott}}} = \frac{n_0}{\varepsilon (1 + \frac{Q^2}{4M^2})} \left[\varepsilon G_E^2 (Q^2) + \frac{Q^2}{4M^2} G_M^2 (Q^2)\right],
\label{Eq:CrossSection}
\end{equation}
where $n_0$ is a normalization factor and the kinematic quantities $Q^2$ and $\varepsilon$ are given by
\begin{align}
Q^2 & = \frac{2M E^2 (1 - \cos{\theta})}{M + E (1 - \cos{\theta})}, \\
\varepsilon & = \left[1 + 2 \left(1 + \frac{Q^2}{4M^2}\right) \tan^2{\frac{\theta}{2}}\right]^{-1}
\end{align}
where $E$ is the incident electron beam energy, $M$~is the proton mass, $\theta$~is the measured electron scattering angle. 
The Mott cross section, $\sigma_{\text{Mott}}$, with the recoil factor included, is given by
\begin{equation}
\sigma_{\text{Mott}}  = \frac{\alpha^2}{4 E^2} \frac{\cos^2{(\theta / 2)}}{\sin^4{(\theta / 2)} ( 1 + \frac{E}{M} (1 - \cos{\theta}))}.
\end{equation}
From Eq.~\ref{Eq:CrossSection} is it clear that as $Q^2$ goes to zero and $\varepsilon$ goes to one 
(forward angle electron scattering), 
the terms which include the  magnetic form factor, $G_M$, become relatively unimportant.

\begin{figure}
\includegraphics[width=\columnwidth]{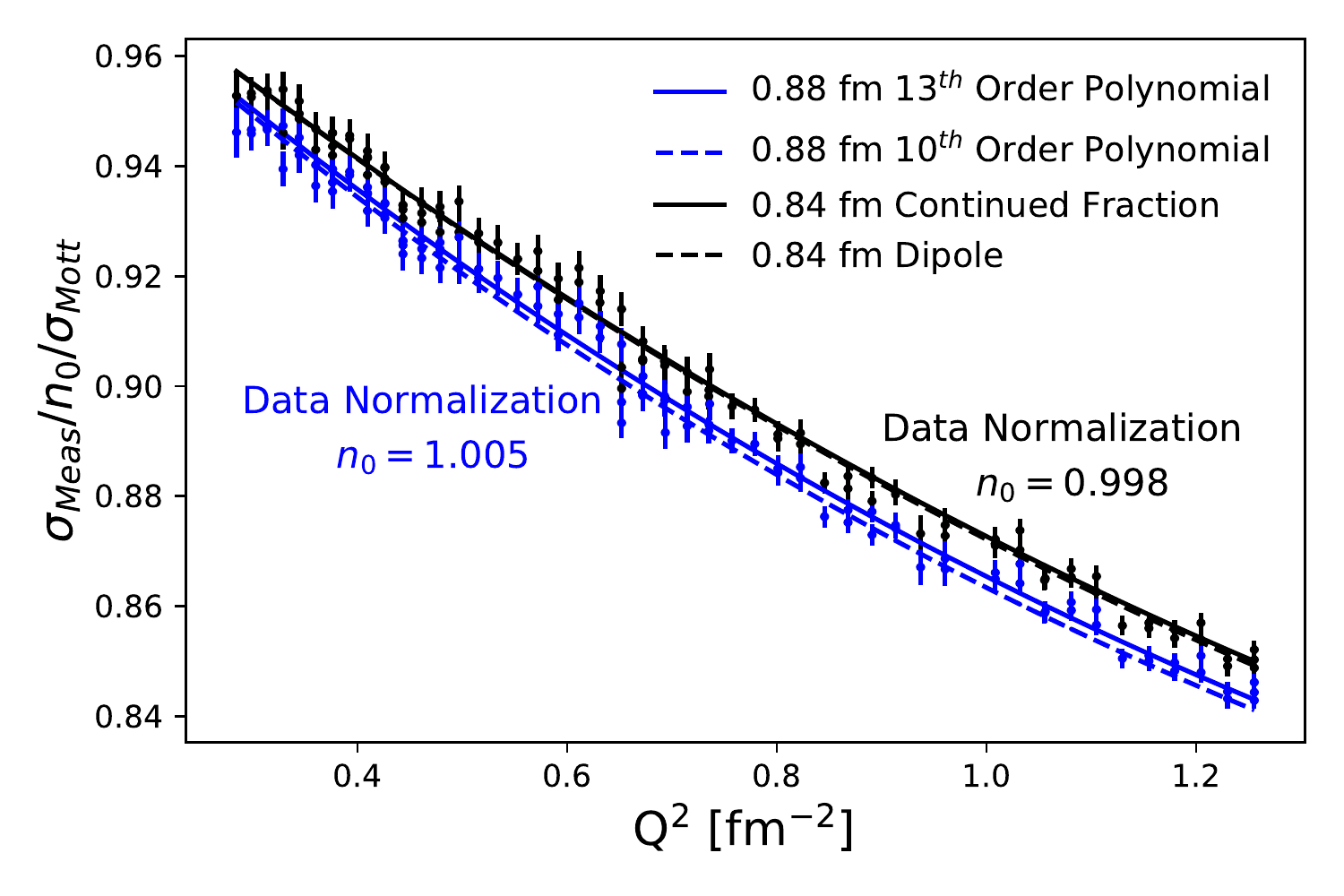} 
\caption{Shown is one set of modern cross section data to illustrate the current tension between 
different models and the effect of the normalization parameter.   In electron scattering data, the
point-to-point uncertainties can be rather small, compared to the typical normalization uncertainty
of a few percents. To make the curves, the standard dipole magnetic
form factor has been used along with charge form factors from several recent publications:  a bounded
13$^{th}$-order polynomial with a radius of 0.88~fm~\cite{Ye:2017gyb}, a unbounded 10$^{th}$-order polynomial with a radius of 0.88~fm~\cite{Bernauer:2013tpr}, 
a continued fraction with a radius of 0.84~fm~\cite{Griffioen:2015hta}, and a dipole function with a radius of 0.84~fm~\cite{Higinbotham:2015rja}. }
\label{RealData}
\end{figure}

To illustrate the current tension between fits done with different models, Fig.~\ref{RealData} shows 
104 data points from one full subset from a modern electron scattering 
experiment~\footnote{Data from Mainz spectrometer B with a beam energy of 315 MeV with radiative and Coulomb corrections~\cite{Bernauer:2013tpr}.} 
that covers a range similar to the range studied herein.  By using a single set, only one floating normalization
parameter is required.
This one set of data covers a range similar to the 85 data point fits with 6 floating normalizations~\cite{Rosenfelder:1999cd,Hill:2010yb}.
Along with the data, four representative functions are shown using Eq.~\ref{Eq:CrossSection} with a standard magnetic form factor.
Two functions give radii that agree with the CODATA value for the proton
radius~\cite{Bernauer:2013tpr,Ye:2017gyb} and two that agree with the muonic Lamb shift measurements~\cite{Higinbotham:2015rja,Griffioen:2015hta}.
We note that this figure looks oddly similar to the pseudo data in Fig.~\ref{linearVSquadratic} but here the true function is unknown 
so it is not clear which curves are shifted with respect to the true reduced cross section values.

Following the logic of this work, we try fitting the cross section data shown in Fig.~\ref{RealData} with  Eq.~\ref{Eq:CrossSection}, where
$G_E$ has been approximated by Eq.~\ref{Eq:linear} and Eq.~\ref{Eq:quadratic}, though with the normalization now subsumed in Eq.~\ref{Eq:CrossSection}.    
Additionally, we performed fits with the following two commonly used functions:
\begin{align}
f_{\mathrm{cubic}}(Q^2)   & = 1 + a_1 Q^2 + a_2 Q^4 + a_3 Q^6,  \\
f_{\mathrm{rational}}(Q^2) & = \frac{1+n_1 Q^2}{1+m_1 Q^2},
\end{align}
where the radius for the rational function is given by $\sqrt{-6 (n_1 - m_1)}$ and for the polynomials by $\sqrt{-6 a_1}$.
The low order rational function ($n=m=1$) can easily be extended to give the expected asymptotic behavior
as $Q^2 \to \infty$ by using a more complex rational function where $m=n+2$ such 
as in~\cite{Kelly:2004hm,Puckett:2017flj,Gutsche:2017lyu}.

In order to check on the influence of the magnetic form factor on the results, 
several different magnetic form factor functions were used.
First, the fits were done with a standard dipole magnetic form factor
and then repeated with the magnetic form factor from~\cite{Bernauer:2013tpr} and~\cite{Ye:2017gyb}.    
The $F$-test, AIC, or BIC model selection techniques all slightly prefer fits with Eq.~\ref{Eq:quadratic}.
The use of statistical criteria for model selection helps avoids confirmation bias,
though one could still be using an inappropriate function for the problem at hand. 
Uncertainties were determined by applying a statistical bootstrap to the data~\cite{Efron:1979}.   This is done
by repeatedly randomly sampling the true data with replacement to generate thousands of new sets of N points 
and refitting those new sets.   This allows one to get uncertainty distributions using the data itself and
avoids a number of assumptions that are required for $\chi^2$ uncertainty techniques 
to be valid~\cite{Efron:1979} and, unlike $\chi^2$ techniques, is also sensitive to over-fitting~\cite{Andrae:2010}.

\begin{table*}
\caption{Using four different magnetic form factor parameterizations, we extract the normalization
and electric radius using a linear, quadratic, cubic, and rational approximation for the $G_E$ function in 
Eq.~\ref{Eq:CrossSection}.
To avoid multiple floating normalizations, the single set of 104  data points shown in Fig.~\ref{RealData} is used.
The parameter uncertainties were obtained  by performing statistical bootstraps of the data.
As seen during the Monte Carlo studies, the linear fit over this interval produces a small variance; 
but is clearly biased from the true 0.84--0.88~fm proton radius, whereas the quadratic fit has a 
larger variance but gives less biased result.   
The cubic fit function over-fits and thus, produces a huge variance.   
The rational function is nearly as good as the quadratic through
produces a systematically larger radius though nicely in the range we expect. In the Table df stands for degrees of freedom.}
\begin{tabular}{cc|cccccl|cccccl}                                                  \hline \hline
\multicolumn{14}{l}{Standard Magnetic Form Factor}                                 \\ \hline
          &     & \multicolumn{6}{c}{without Coulomb correction}                 & \multicolumn{6}{|c}{with Coulomb correction} \\
$G_E$ Fit & df  & $\chi^2$ & reduced   & AIC    & BIC    & Norm      & Extrapolated & $\chi^2$ & reduced   & AIC    & BIC    & Norm      & Extrapolated      \\  
Function  &     &          & $\chi^2$  &        &        &           & Radius [fm]      &          & $\chi^2$  &        &        &           & Radius [fm]        \\ \hline
linear    & 102 & 162.6    & 1.594     & 50.47  & 55.74  & 0.988(1)  & 0.785(2)  & 167.2    & 1.639     & 53.35  & 58.64  & 0.985(1)  & 0.790(2)    \\
quadratic & 101 & 119.4    & 1.182     & 20.35  & 28.29  & 1.000(2)  & 0.852(10) & 119.0    & 1.178     & 20.00  & 27.93  & 0.998(2)  & 0.860(10)   \\
cubic     & 100 & 117.3    & 1.173     & 20.51  & 31.13  & 0.990(6)  & 0.785(57) & 117.1    & 1.171     & 20.33  & 30.90  & 0.990(6)  & 0.797(57)   \\    
rational  & 101 & 120.0    & 1.188     & 20.83  & 28.76  & 1.001(2)  & 0.860(10) & 119.6    & 1.184     & 20.50  & 28.64  & 0.999(2)  & 0.869(10)   \\ \hline \hline
\multicolumn{14}{l}{Kelly Magnetic Form Factor}                                   \\ \hline
          &     & \multicolumn{6}{c}{without Coulomb correction}                 & \multicolumn{6}{|c}{with Coulomb correction} \\
$G_E$ Fit & df  & $\chi^2$ & reduced   & AIC    & BIC    & Norm      & Extrapolated    & $\chi^2$ & reduced   & AIC    & BIC    & Norm      & Extrapolated     \\  
Function  &     &          & $\chi^2$  &        &        &           & Radius [fm]      &          & $\chi^2$  &        &        &           & Radius [fm]            \\ \hline
linear    & 102 & 168.1    & 1.648     & 53.95  & 59.24  & 0.986(1)  & 0.780(2)  & 173.0    & 1.669     & 53.35  & 58.64  & 0.985(1)  & 0.785(2)    \\
quadratic & 101 & 119.4    & 1.182     & 20.33  & 28.27  & 1.000(2)  & 0.852(10) & 119.0    & 1.178     & 20.00  & 27.91  & 0.998(2)  & 0.860(10)   \\
cubic     & 100 & 117.1    & 1.171     & 20.36  & 30.94  & 0.985(6)  & 0.598(57) & 116.9    & 1.169     & 20.14  & 30.72  & 0.983(6)  & 0.613(57)   \\    
rational  & 101 & 120.0    & 1.188     & 20.89  & 28.82  & 1.001(2)  & 0.861(10) & 119.6    & 1.188     & 20.87  & 28.80  & 1.001(2)  & 0.870(10)   \\ \hline \hline
\multicolumn{14}{l}{Bernauer Magnetic Form Factor}                                 \\ \hline
          &     & \multicolumn{6}{c}{without Coulomb correction}                 & \multicolumn{6}{|c}{with Coulomb correction} \\
$G_E$ Fit & df  & $\chi^2$ & reduced   & AIC    & BIC    & Norm      & Extrapolated    & $\chi^2$ & reduced   & AIC    & BIC    & Norm      & Extrapolated      \\  
Function  &     &          &$\chi^2$  &        &       &             & Radius [fm]      &          & $\chi^2$  &        &        &           & Radius [fm]            \\ \hline
linear    & 102 & 163.3    & 1.601     & 53.96  & 56.24 & 0.988(1)   & 0.786(2)  & 168.0    & 1.647     & 53.85  & 59.14  & 0.985(1)  & 0.791(2)    \\
quadratic & 101 & 119.5    & 1.183     & 20.44  & 28.37 & 1.000(2)   & 0.854(10) & 119.0    & 1.179     & 20.08  & 28.00  & 1.000(4)  & 0.862(10)    \\ 
cubic     & 100 & 117.3    & 1.173     & 20.52  & 31.13 & 0.992(6)   & 0.786(56) & 117.1    & 1.171     & 20.33  & 30.91  & 0.990(6)  & 0.797(56)    \\ 
rational  & 102 & 120.0    & 1.189     & 20.93  & 28.86 & 1.001(2)   & 0.871(13) & 119.7    & 1.185     & 20.60  & 28.54  & 0.999(2)  & 0.871(13)    \\ \hline \hline
\multicolumn{14}{l}{Ye Magnetic Form Factor}                                      \\ \hline
          &     & \multicolumn{6}{c}{without Coulomb correction}                 & \multicolumn{6}{|c}{with Coulomb correction} \\
$G_E$ Fit & df  & $\chi^2$ & reduced   & AIC    & BIC    & Norm      & Extrapolated    & $\chi^2$ & reduced   & AIC    & BIC    & Norm      & Extrapolated      \\  
Function  &     &          &$\chi^2$  &        &       &             & Radius [fm]      &          & $\chi^2$  &        &        &           & Radius [fm]    \\ \hline
linear    & 102 & 167.2    & 1.639    & 53.35  & 58.64 & 0.987(1)    & 0.781(2)  & 172.0    & 1.686     & 56.33  & 61.62  & 0.984(1)  & 0.786(2)    \\
quadratic & 101 & 119.4    & 1.182    & 20.33  & 28.26 & 1.000(2)    & 0.852(10) & 119.0    & 1.190     & 19.97  & 27.91  & 0.998(2)  & 0.860(10)    \\
cubic     & 100 & 117.3    & 1.173    & 20.55  & 31.12 & 0.992(6)    & 0.785(55) & 117.1    & 1.171     & 20.33  & 30.91  & 0.992(6)  & 0.797(55)    \\ 
rational  & 102 & 119.8    & 1.188    & 20.87  & 28.80 & 1.001(2)    & 0.860(14) & 119.6    & 1.184     & 20.54  & 28.48  & 0.999(2)  & 0.870(14)    \\ \hline \hline
\end{tabular}
\label{datatable}
\end{table*}

It is of particular note that the quadratic fit is the same function over a similar 
range as found in the original 1976 work opting for
the quadratic fit~\cite{Borkowski:1975ume} though herein we 
use a single floating normalization instead of three and the point-to-point errors are smaller.   
It is perhaps distressing that published values of the 
radius extracted from electron scattering remained basically unchanged since the 1976 work~\cite{Borkowski:1975ume} while the 
functions used to make the extrapolation and obtain that same radius became increasingly complex and convoluted.
Oddly enough the standard dipole magnetic form factor quadratic fit has the lowest AIC and BIC values and gives a result consistent
with the muonic Lamb shift; though the rational function is nearly as good and is nearly exactly between the CODATA
and muonic Lamb shift values.

\section{Graphs in Statistical Analysis}

Beyond simply checking statistic criteria, it is important to check the statistical analysis graphically~\cite{Anscombe:1973}.
The graphs help ensure that our underlying assumptions about the data are reasonably correct and help establish that the results
aren't being overly influenced by a single point.   This is particularly important when doing $\chi^2$ minimizations where a
clear outlier can easily have an undue weight, given the quadratic structure of $\chi^2$ in the errors. This is not to imply that one should simply remove an outlier, but it does
mean that one might wish to study the effect of fitting with and without the outlining point to clearly show its influence on 
the result.   The researcher can also review their notes to ensure that nothing odd happened during the taking of that data point.
To illustrate these points clearly, one can find in Appendix~\ref{appendix-quartet} an updated version of the classic Anscombe example~\cite{Anscombe:1973} 
where we have added uncertainties so all the examples give a reduced $\chi^2$ of unity yet the distribution of the data are in
fact all clearly different.

\begin{figure}[htb]
\includegraphics[width=\columnwidth]{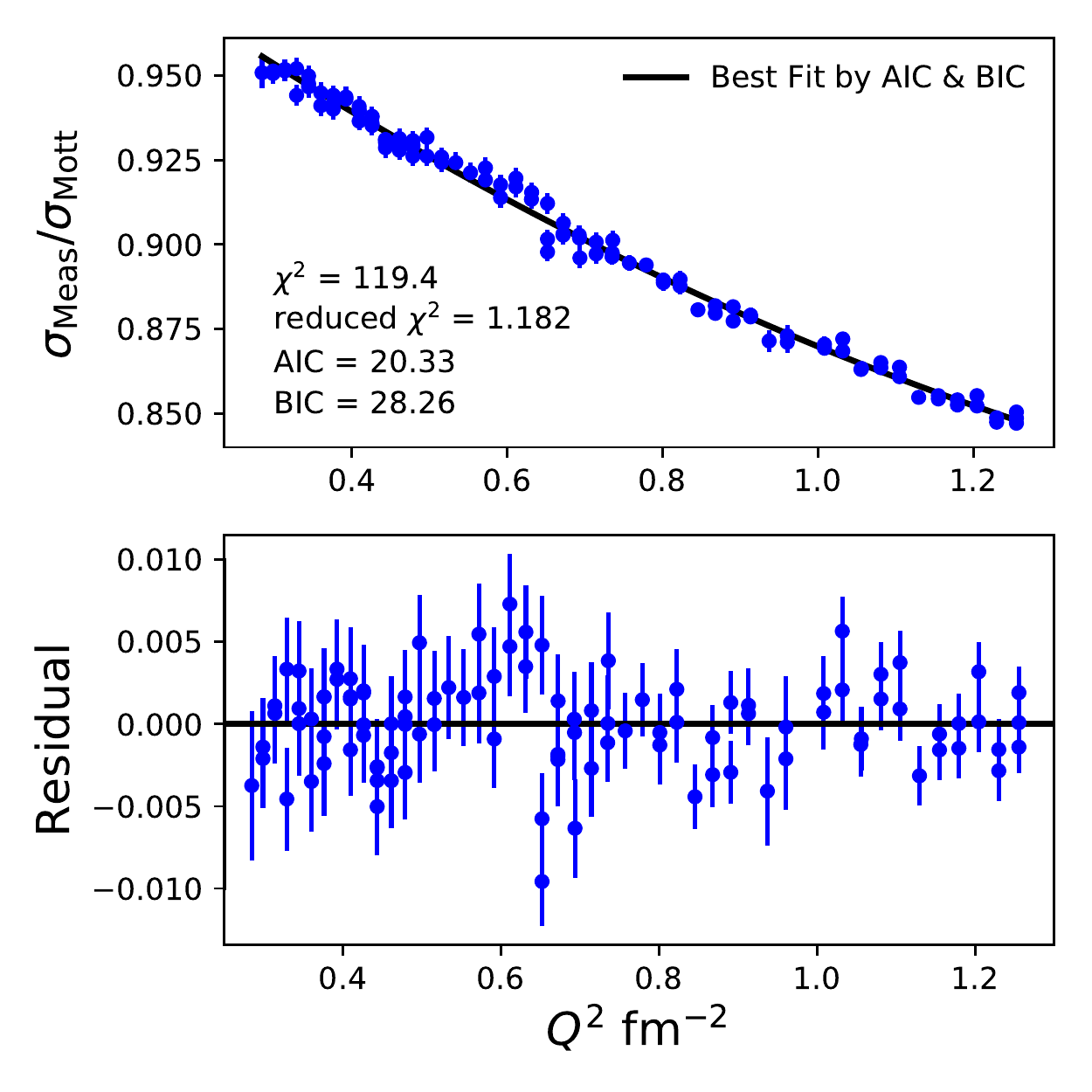}
\caption{The cross section data and the quadratic fit curve are shown in the upper plot and the residual, the difference between
the data and fit, are shown in the lower plot.    If an appropriate fit function was used, the residuals should be randomly
dispersed around the horizontal axis.}
\label{residual}
\end{figure}

\begin{figure}[htb]
\includegraphics[width=\columnwidth]{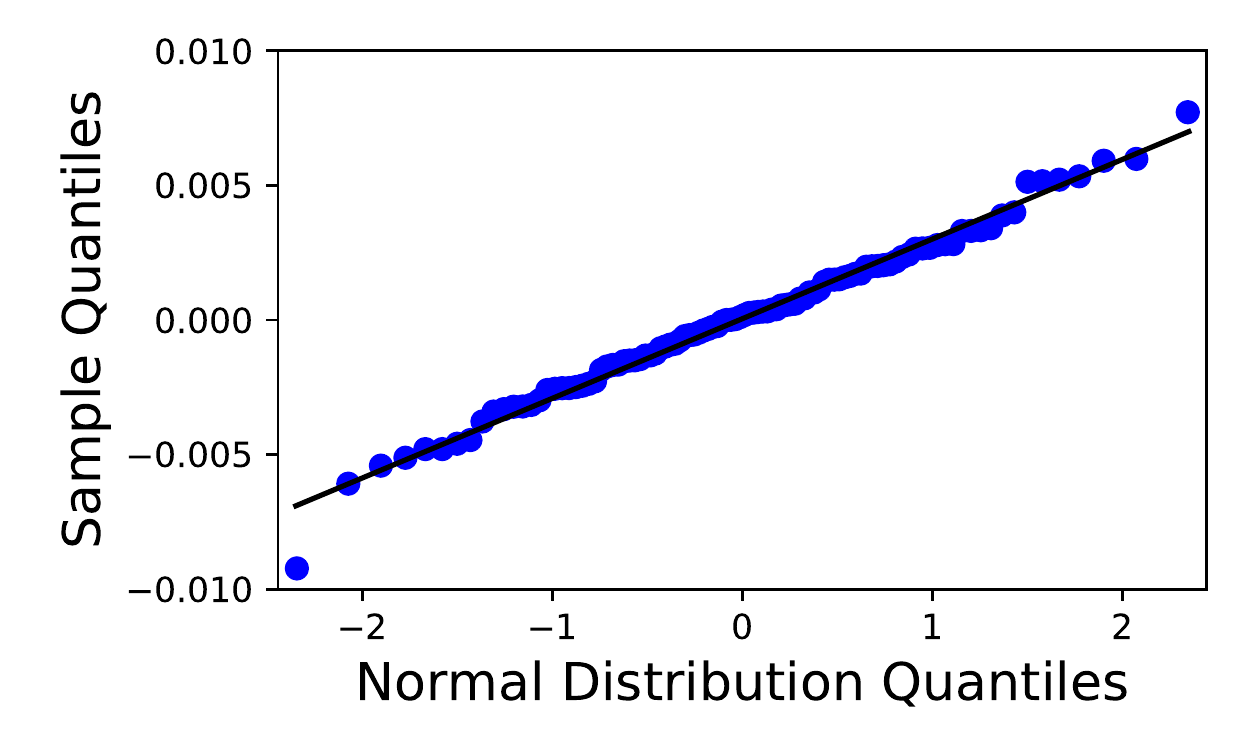}
\caption{In order to check if the distribution of the residuals are plausibly normally distributed, one can make a normal
 quantile-quantile, Q-Q, plot.
This is done by sorting the residuals from lowest to highest and plotting the ordered residuals against a theoretical 
normal distribution.   The resulting plot is a beautiful example of normally distributed residuals.}
\label{normqq}
\end{figure}

In the top half of Fig.~\ref{residual}, the reported data is shown along with the quadratic fit:  best fit from Table~\ref{datatable} based
on AIC and BIC.   In the bottom half, the residual, a difference between the data and the model, is shown.     While these two plots
are perhaps the most common statistical analysis graphs, they are not found in many cited proton radius papers.  In fact, some
papers have no data quality plots at all (e.g.~\cite{Rosenfelder:1999cd}).
As the human eye will find patterns in statistical noise, the normal Q-Q plot is an important tool to check whether or not the 
data is normally distributed~\cite{Wilk:1968}.   This is done in Fig.~\ref{normqq}, where the sorted residuals are plotted against,
and are shown to follow, a normal distribution.   There are of course more formal statistical tests one can apply, but in general the
visualizations of the data can quickly reveal if there are any major problems.   In particular, with least squares fitting, one should
be mindful that a single outlier can skew the results.

\section{Lowest $Q^2$ Data} 

\begin{figure}[htb]
\includegraphics[width=\columnwidth]{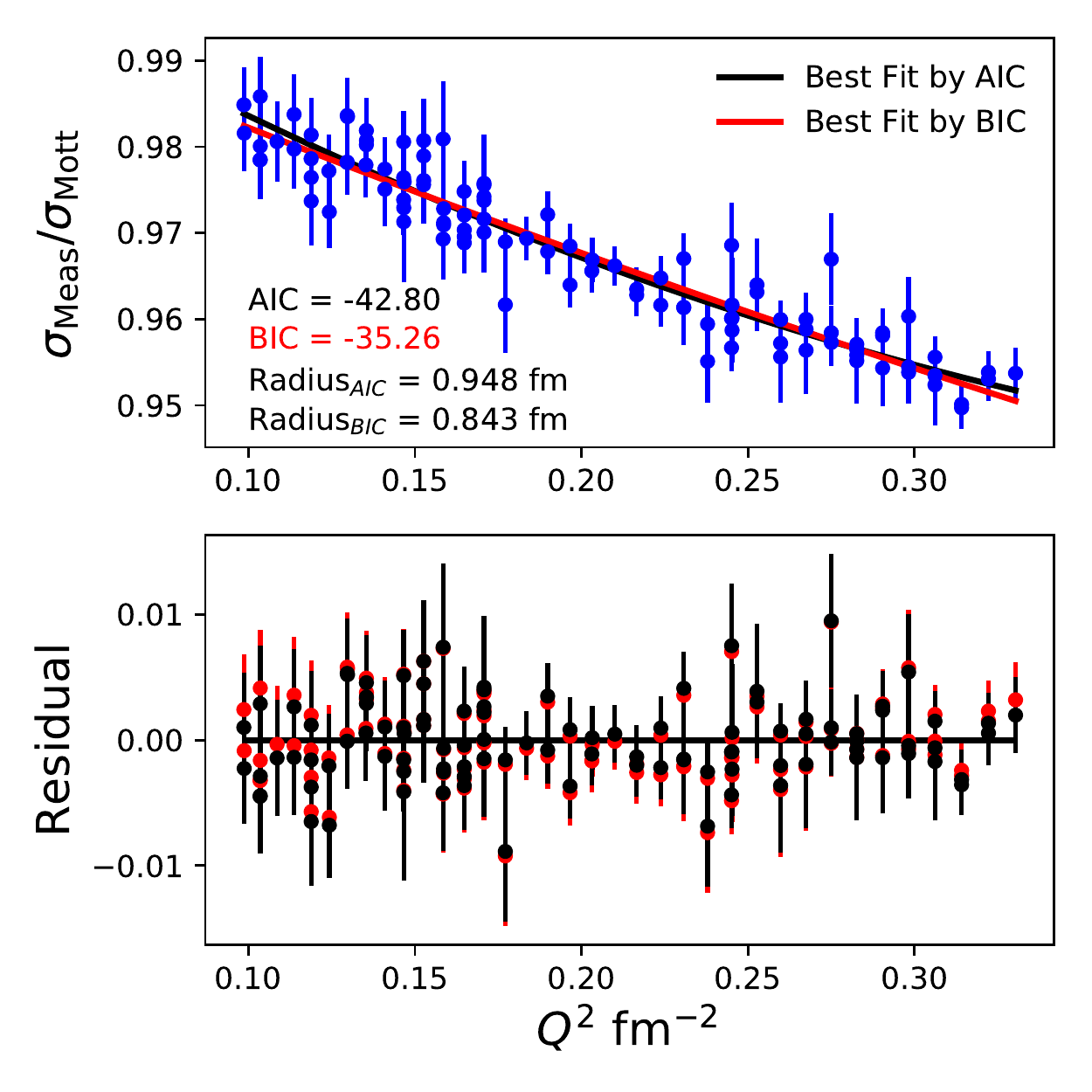}
\caption{The cross section data and the quadratic fit curve are shown in the upper plot and the residual, the difference between
the data and fit, are shown in the lower plot.}
\label{residual-lowest}
\end{figure}

\begin{figure}[htb]
\includegraphics[width=\columnwidth]{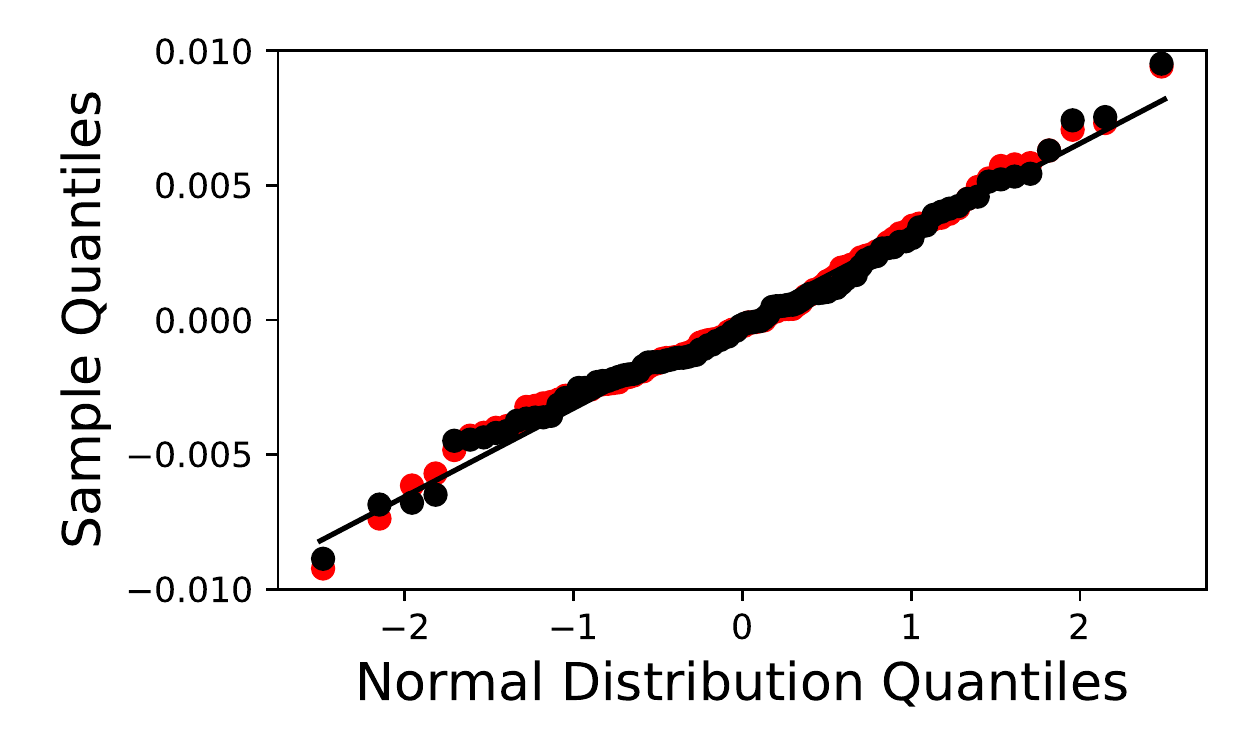}
\caption{
The low $Q^2$ data quite as normally distributed residuals though still reasonable close to what one would expect.}
\label{normqq-lowest}
\end{figure}

While the set of Mainz data studied above had a range similar to the large range of our Monte Carlo studies, it is not
the lowest $Q^2$ set of data. The lowest set comes from a run with 180 MeV~\cite{Higinbotham:2019jzd}
and covers a range from 0.090 to 0.330 fm$^{-2}$.   This range is particularly intriguing as it has been shown to be
low enough that even a linear regression should be able to extract the radius; though instead of just relying on the Monte
Carlo study to make that choice, we once again systematically look at the data with various function and variance choices
for the magnetic form factor in Table~\ref{lowestdatatable}.   The residuals and normal Q-Q plots 
for the AIC and BIC selected
models are shown in Fig.~\ref{residual-lowest} and \ref{normqq-lowest} respectively.

\begin{table*}
\caption{Using four different magnetic form factor parameterizations, we extract the normalization
and electric radius using a linear, quadratic, cubic, and rational approximation for the $G_E$ function in 
Eq.~\ref{Eq:CrossSection}.
To avoid multiple floating normalizations, the single set of 106  data points 
shown in Fig.~\ref{RealData} is used.
The parameter uncertainties were obtained by performing statistical bootstraps of the data.
As seen during the Monte Carlo studies, the linear fit over this interval produces a small variance; 
but is clearly biased from the true 0.84--0.88~fm proton radius, whereas the quadratic has a 
larger variance but gives less biased result.   The cubic fit function is over-fitting which is 
why it produces a huge variance.   The rational function is nearly as good as the quadratic through
produces a systematically larger radius though nicely in the range we expect. In the Table df stands for degrees of freedom.}
\begin{tabular}{cc|cccccl|cccccl}                                                  \hline \hline
\multicolumn{14}{l}{Standard Magnetic Form Factor}                                 \\ \hline
          &     & \multicolumn{6}{c}{without Coulomb correction}                 & \multicolumn{6}{|c}{with Coulomb correction} \\
$G_E$ Fit & df  & $\chi^2$ & reduced   & AIC    & BIC    & Norm      & Extrapolated& $\chi^2$ & reduced   & AIC    & BIC    & Norm      & Extrapolated    \\  
Function  &     &          & $\chi^2$  &        &        &           & Radius [fm]      &          & $\chi^2$  &        &        &           & Radius [fm]        \\ \hline
linear    & 104 &  69.2    & 0.666     &$-41.14$&$-35.81$& 1.000(1)  & 0.826(8)  &  69.6    & 0.670     &$-40.53$&$-35.21$& 0.997(1)  & 0.842(8)    \\
quadratic & 103 &  67.0    & 0.651     &$-42.58$&$-34.59$& 1.006(3)  & 0.923(50) &  66.9    & 0.649     &$-42.80$&$-34.81$& 1.004(3)  & 0.948(50)   \\
cubic     & 102 &  66.3    & 0.650     &$-41.79$&$-31.13$& 0.996(6)  & 0.598(nan)&  66.2    & 0.649     &$-41.95$&$-31.30$& 0.994(6)  & 0.645(nan)   \\    
rational  & 103 &  63.8    & 0.632     &$-43.77$&$-30.46$& 0.934(3)  & 0.936(55) &  67.0    & 0.650     &$-42.63$&$-34.64$& 1.004(3)  & 0.964(55)   \\ \hline \hline
\multicolumn{14}{l}{Kelly Magnetic Form Factor}                                   \\ \hline
          &     & \multicolumn{6}{c}{without Coulomb correction}                 & \multicolumn{6}{|c}{with Coulomb correction} \\
$G_E$ Fit & df  & $\chi^2$ & reduced   & AIC    & BIC    & Norm      & Extrapolated    & $\chi^2$ & reduced   & AIC    & BIC    & Norm      & Extrapolated      \\  
Function  &     &          & $\chi^2$  &        &        &           & Radius [fm]      &          & $\chi^2$  &        &        &           & Radius [fm]            \\ \hline
linear    & 104 &  69.3    & 0.666     &$-41.03$&$-35.70$& 1.000(1)  & 0.824(8)  &  69.7    & 0.670     &$-40.41$&$-35.08$& 0.998(1)  & 0.840(8)    \\
quadratic & 103 &  67.0    & 0.651     &$-42.58$&$-34.59$& 1.000(2)  & 0.923(50) &  66.9    & 0.649     &$-42.80$&$-34.81$& 1.004(2)  & 0.948(50)   \\
cubic     & 102 &  66.2    & 0.649     &$-41.86$&$-31.21$& 0.995(6)  & 0.525(nan)&  66.1    & 0.648     &$-42.03$&$-31.38$& 0.993(6)  & 0.578(nan)   \\    
rational  & 103 &  67.1    & 0.652     &$-42.43$&$-34.44$& 1.006(3)  & 0.937(55) &  67.0    & 0.651     &$-42.62$&$-34.63$& 1.005(3)  & 0.965(55)   \\ \hline \hline
\multicolumn{14}{l}{Bernauer Magnetic Form Factor}                                 \\ \hline
          &     & \multicolumn{6}{c}{without Coulomb correction}                 & \multicolumn{6}{|c}{with Coulomb correction} \\
$G_E$ Fit & df  & $\chi^2$ & reduced   & AIC    & BIC    & Norm      & Extrapolated  & $\chi^2$ & reduced   & AIC    & BIC    & Norm      & Extrapolated      \\  
Function  &     &          &$\chi^2$   &        &        &           & Radius [fm]   &          & $\chi^2$  &        &        &           & Radius [fm]            \\ \hline
linear    & 104 &  69.2    & 0.665     &$-41.19$&$-35.87$& 1.001(1)  & 0.826(8)  &  69.9    & 0.669     &$-40.59$&$-35.26$& 0.998(1)  & 0.842(8)    \\
quadratic & 103 &  67.0    & 0.651     &$-42.58$&$-34.59$& 1.006(2)  & 0.923(50) &  66.9    & 0.649     &$-42.80$&$-34.81$& 1.004(4)  & 0.948(50)    \\ 
cubic     & 102 &  66.2    & 0.650     &$-41.79$&$-31.13$& 0.996(6)  & 0.598(nan)&  66.2    & 0.649     &$-41.95$&$-31.30$& 0.994(6)  & 0.645(nan)    \\ 
rational  & 103 &  67.1    & 0.652     &$-42.44$&$-34.44$& 1.007(3)  & 0.936(55) &  67.0    & 0.650     &$-42.63$&$-34.64$& 0.999(3)  & 0.964(55)    \\ \hline \hline
\multicolumn{14}{l}{Ye Magnetic Form Factor}                                      \\ \hline
          &     & \multicolumn{6}{c}{without Coulomb correction}                 & \multicolumn{6}{|c}{with Coulomb correction} \\
$G_E$ Fit & df  & $\chi^2$ & reduced   & AIC    & BIC    & Norm      & Extrapolated & $\chi^2$ & reduced   & AIC    & BIC    & Norm      & Extrapolated      \\  
Function  &     &          &$\chi^2$   &        &        &           & Radius [fm]      &          & $\chi^2$  &        &        &       & Radius [fm]     \\ \hline
linear    & 104 &  69.3    & 0.666     &$-41.04$&$-35.71$& 1.000(1)  & 0.824(8)  &  69.7    & 0.670     &$-40.42$&$-35.09$& 0.998(1)  & 0.841(8)    \\
quadratic & 103 &  67.0    & 0.651     &$-42.58$&$-34.59$& 1.006(2)  & 0.923(50) &  66.9    & 0.649     &$-42.80$&$-34.81$& 1.004(2)  & 0.948(50)    \\
cubic     & 102 &  66.3    & 0.650     &$-41.79$&$-31.13$& 0.996(6)  & 0.598(nan)&  66.2    & 0.649     &$-41.95$&$-31.30$& 0.994(6)  & 0.645(nan)    \\ 
rational  & 103 &  67.1    & 0.652     &$-42.43$&$-34.44$& 1.007(3)  & 0.936(55) &  67.0    & 0.651     &$-42.62$&$-34.63$& 1.005(3)  & 0.964(55)    \\ \hline \hline
\end{tabular}
\label{lowestdatatable}
\end{table*}

Although these fits look beautiful, the extraction of the radius depends on
our belief that we can reasonably expect our function to extrapolate well.
This belief is based on the Monte Carlo study at the beginning of the paper.    And while it is certainly reasonable
to assume the charge form factor will smoothly continue to $Q^2$ = 0,
nature will do as it wishes. New data in the less then 0.1~fm$^{-2}$ region  is 
certainly desirable to ensure that our anzats is correct~\cite{Gasparian:2014rna,
Peng:2016szv, Mihovilovic:2016rkr,Yan:2018bez}.   
Also, strictly speaking, the
uncertainties from the regressions are only valid over the range of the data, so again our 
belief that we are obtaining reasonable uncertainties links back to the Monte Carlo study of the
radius extraction using different functions and the relatively short distance of the extrapolation.

As has been shown by theory calculations
such as Alarcon and Weiss~\cite{Alarcon:2017lhg,Alarcon:2018irp}
the moments of the generating function are likely far too complex to be constrained
by descriptive fitting of experimental data.  Also, as pointed out on page 378 of \cite{Sirca:2016}, 
moments do not uniquely define functions, so the best
we can do for the higher order terms is determine the shape of the data and then compare with theory.

One can of course try to combine multiple sets of data, though this quickly turns into a Bayesian exercise
with no unique solution and has an inherent human-in-the-loop factor~\cite{Daee:2018:UMA:3172944.3172989}.
One could also try using a physical model and just simply compare the model with a descriptive fit of the 
data~\cite{Alarcon:2018zbz, Alarcon:2018irp,Alarcon:2017ivh,Alarcon:2017chi,Alarcon:2017lhg,Alarcon:2012kn}.

In addition, in Appendix~\ref{appendix-stepwise} we show an example of using a machine learning significance testing technique, forward stepwise regression, to find the best polynomial fit function.   
And in Appendix~\ref{appendix-mapping} we also show
the effect of conformal mapping in order to clearly show that even after doing a mapping 
one will still need to apply a rigorous model selection criteria.

\section{Summary}

This work was not meant to be an exhaustive study of the world electron scattering data; but 
instead an exploration of bias-variance trade-off as it relates to the extraction of the proton 
radius from electron scattering data.
To do this, we have revisited a classic Monte Carlo study~\cite{Borkowski:1975ume} and shown that
simply rejecting models with a bias is incorrect.
Although adding more regression parameters usually reduces the bias, it comes at the cost of increased variance.
We also illustrated how parsimonious models can have better 
predictive power than even the true underlying model in certain situations (see also section~\ref{SemiAnalytical} for semi analytical calculations on this regard). 

Next, we carefully studied high precision sets of Mainz low $Q^2$ data~\cite{Bernauer:2010wm} 
which covered a range very similar to the Monte Carlo study.   Here we have defined a set to be a
group of data with a single normalization parameter.  The model selection techniques
presented herein provide a rigorous method of selecting an appropriate model to describe a given set of data.  
We have illustrated that these model selection techniques along with some key statistical 
analysis plots can help to ensure a reasonable fit.   

These ideas and techniques of model selection are not limited to the physical sciences, but also extend
to quantitative analysis~\cite{Brighton:2015}, and sit at the heart of statistical 
learning~\cite{Hastie:2009}.  It has been argued that over-fitting is perhaps more problematic now than
in the past due modern computing power~\cite{Cawley:2010}.   It is amusing to note it
was just 1985 when Feynman noted that computers could not beat humans at the
game of Go~\cite{Feynman:2008}, while today computers dominate even this complicated game~\cite{Silver:2016,Barradas:2018}.
With all this computer power available, it is perhaps more necessary then ever
to keep in mind the power and importance of parsimonious modeling as nicely 
summarized by the renowned statistician George Box: 
``Since all models are wrong the scientist cannot obtain a `correct' one
by excessive elaboration.  On the contrary, following William of Occam, 
[the scientist] should seek an economical description of natural phenomena. 
Just as the ability to devise simple but evocative models is the signature of the
great scientist so over-elaboration and over-parameterization is often
the mark of mediocrity.''~\cite{Box76}

\begin{acknowledgments}
We would like to thank Miha~Mihovilovi\v{c} for many extremely useful discussions
and help developing a number of the figures.
The regressions done in the body of this work were done in Python making use of the
outstanding LMFIT package~\cite{Newville:2014} to interface with SciPy
libraries~\cite{Jones:2001}.  Non-linear regressions were done using the
Levenberg–Marquardt method~\cite{Levenberg:1944,Marquardt:1963}.  Special
thanks to Edward Tufte and his efforts to help improve the visualization of 
evidence~\cite{Tufte:1986,Tufte:1990,Tufte:1997,Tufte:2006}.
This work was supported by the U.S.  Department of Energy contract DE-AC05-06OR23177
under which Jefferson Science Associates operates the Thomas Jefferson National 
Accelerator Facility and contract DE-FG02-03ER4123 at Duke University.
\end{acknowledgments}

\begin{appendix}

\section{Anscombe's Quartet}
\label{appendix-quartet}

With the power of modern computing, one can be tempted to blindly assume the results of
a calculation are correct; but this can be extremely misleading~\cite{Sirca:2012}.  To illustrate
this point we use the Anscombe quartet~\cite{Anscombe:1973}, though as nuclear physicists tend 
to use reduced $\chi^2$ instead of R$^2$, we have taken the 1973 example
problem and added uncertainties to the points as shown in Table~\ref{quartet}.

\begin{table}[htb]
\caption{Four data sets of (x,y,dy) values.}
\begin{tabular}{rccccccc}
\multicolumn{1}{l}{Data set}    & 1-3  & 1    & 2    & 3    & 4   & 4    & 1-4   \\
\multicolumn{1}{l}{Variable}    & x    & y    & y    & y    & x   & y    & dy    \\  
                                &      &      &      &      &     &      &       \\
Obs. no. 1: & 10.0 & 8.04 & 9.14 & 7.46 & 8.0 & 6.58 & 1.235 \\ 
         2: &  8.0 & 6.95 & 8.14 & 6.77 & 8.0 & 5.76 & 1.235 \\
         3: & 13.0 & 7.58 & 8.74 &12.74 & 8.0 & 7.71 & 1.235 \\
         4: &  9.0 & 8.81 & 8.77 & 7.11 & 8.0 & 8.84 & 1.235 \\
         5: & 11.0 & 8.33 & 9.26 & 7.81 & 8.0 & 8.47 & 1.235 \\
         6: & 14.0 & 9.96 & 8.10 & 8.84 & 8.0 & 7.04 & 1.235 \\
         7: &  6.0 & 7.24 & 6.13 & 6.08 & 8.0 & 5.25 & 1.235 \\
         8: &  4.0 & 4.26 & 3.10 & 5.39 &19.0 &12.50 & 1.235 \\
         9: & 12.0 &10.84 & 9.13 & 8.15 & 8.0 & 5.56 & 1.235 \\
        10: &  7.0 & 4.82 & 7.26 & 6.42 & 8.0 & 7.91 & 1.235 \\
        11: &  5.0 & 5.68 & 4.74 & 5.73 & 8.0 & 6.89 & 1.235 \\ 
            &      &      &      &      &     &      &       \\ \hline
\end{tabular}
\label{quartet}
\end{table}

\begin{figure}[htb]
\includegraphics[width=\columnwidth]{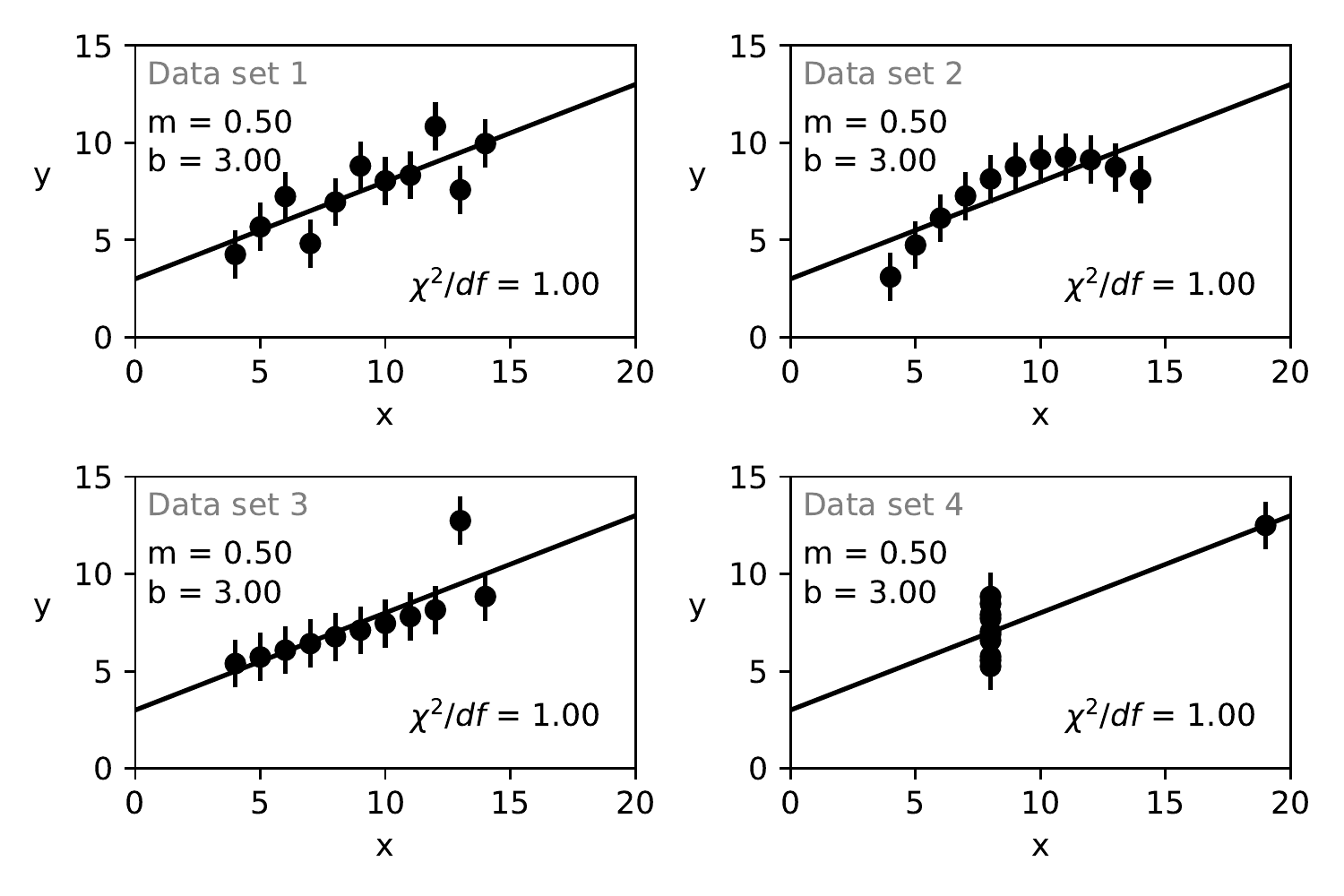}
\caption{Graphs of the four sets of data.}
\label{SameChi2}
\end{figure}

These four sets of (x,y,dy) values give to three significant figures the
same statistical quantities: mean, variance, $\chi^2$, reduced $\chi^2$, etc.
So if one fails to make graphical checks, one can be completely fooled into thinking
the fits are all equally good; but by simply graphing (see Fig.~\ref{SameChi2}) 
one can see that only the first set 
of data is distributed in an ideal way around the fit function.

Data set two clearly has a curved residual yet has exactly the 
same mean, errors, $\chi^2$ as the first fit.   This suggests that the fitter 
should likely add a quadratic term to the regression as well as check that
the uncertainties have been correctly reported.

Data set three illustrates the effect of an outlier on the regression.
Of course, the scientist doesn't simply throw out an outlier.
Instead one should report on the outlier's influence on the result.   For example,
in data set three it would be worth noting that if the outlier is removed, the data
exactly follow a line of $y = 4 + 0.346 x$ and that that measurement should be repeated.

Set four looks unsatisfactory, since all the information about the slope comes
from one observation.   This is very different from data set one where any one point
can be removed and one will obtain nearly the same result.   Thus, for this data set
four it should be pointed out that a single observation plays a critical role
in the result.

\section{Stepwise Regression}
\label{appendix-stepwise}

In this paper, we have proceeded as is natural for a human statistical modeler, starting from a simple
linear approximation and then adding additional terms in $Q^2$ as needed.  Of course nothing
says that this type of fitting determines the terms of the moments the
true generating function but simply that we have found an appropriate function for fitting the data.
In fact, using experimentally determined ``moments'' as a regression constraint can lead to circular logic
as the normalization parameters and moments are linked in a very complex way.

In Ref.~\cite{Higinbotham:2015rja} the authors used forward stepwise regression on the Mainz
Rosenbluth data~\cite{Bernauer:2010wm} and the code returned a beautiful alternating sign power 
series which is what one would expect if the power series fit was in fact extracting moments
of a generating function.    Nevertheless, without a physical theory in mind, no claim
was made that these were the moments of the generating function.

In fact, if one applies the same stepwise regression code to very low $Q^2$ data it can give
some seemly nonsensical results.  While the linear term will remain essentially unchanged one no longer
get the series of alternating .   
As an example, using R~\cite{R} with the CAR package~\cite{car} to preform stepwise regression with AIC as the
model selection criteria on the $G_E$ data from Griffioen, 
Carlson and Maddox~\cite{Griffioen:2015hta} with $Q^2 < 0.8$ fm$^{-2}$, to match a range studied
in the Monte Carlo simulations, one obtains the following result:

\begin{Verbatim}[fontsize=\footnotesize]

Start:  AIC=36.47
data$y ~ data$x

               Df Sum of Sq    RSS    AIC
+ I(data$x^4)   1    9.1204 357.98 30.090
+ I(data$x^3)   1    9.1063 358.00 30.104
+ I(data$x^5)   1    8.9767 358.13 30.224
+ I(data$x^2)   1    8.8697 358.23 30.324
+ I(data$x^6)   1    8.7322 358.37 30.451
+ I(data$x^7)   1    8.4320 358.67 30.730
+ I(data$x^8)   1    8.1084 359.00 31.030
+ I(data$x^9)   1    7.7822 359.32 31.333
+ I(data$x^10)  1    7.4655 359.64 31.626
+ I(data$x^11)  1    7.1643 359.94 31.905
<none>                      367.10 36.468

Step:  AIC=30.09
data$y ~ data$x + I(data$x^4)

               Df Sum of Sq    RSS    AIC
<none>                      357.98 30.090
+ I(data$x^2)   1  0.028238 357.96 32.064
+ I(data$x^8)   1  0.021000 357.96 32.071
+ I(data$x^7)   1  0.020418 357.96 32.071
+ I(data$x^9)   1  0.020399 357.96 32.071
+ I(data$x^3)   1  0.018965 357.96 32.073
+ I(data$x^6)   1  0.018854 357.96 32.073
+ I(data$x^10)  1  0.018801 357.96 32.073
+ I(data$x^5)   1  0.017067 357.97 32.075
+ I(data$x^11)  1  0.016573 357.97 32.075

Weighted Residuals:
     Min       1Q   Median       3Q      Max 
-3.01533 -0.73635 -0.08819  0.66435  3.08064 

Coefficients:
              Estimate Std. Error t value Pr(>|t|)    
(Intercept)  0.9988295  0.0003569  2798.9  < 2e-16 
data$x      -0.1172160  0.0011112  -105.5  < 2e-16 
I(data$x^4)  0.0061991  0.0021379     2.9  0.00399 

Residual standard error: 1.042 on 330 degrees of freedom
Multiple R-squared:  0.9931,  Adjusted R-squared:  0.993 
F-statistic: 2.37e+04 on 2 and 330 DF, p-value: < 2.2e-16

\end{Verbatim}

\begin{figure}[htb]
\includegraphics[width=\columnwidth]{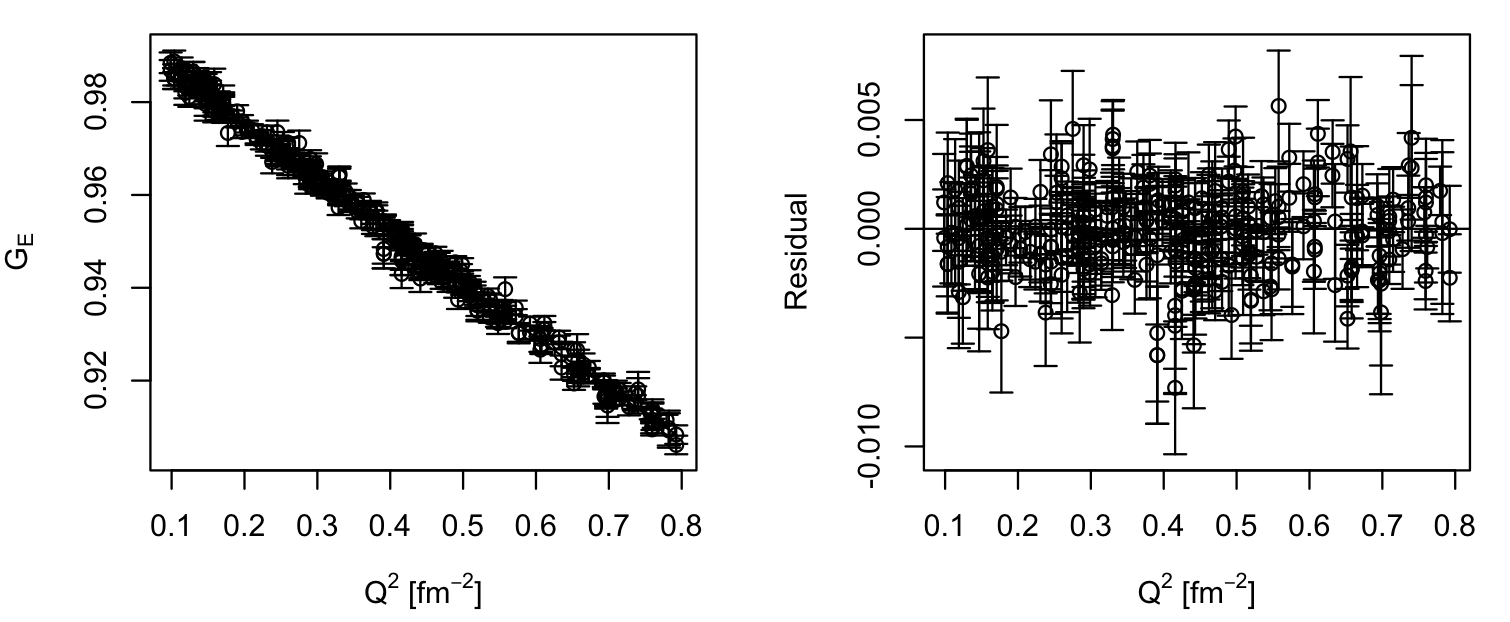}
\caption{The $G_E$ data of~\cite{Griffioen:2015hta} shown along with a residual of the best fit using forward stepwise regression.}
\label{R-stepwise}
\end{figure}

Unlike the previous studies in this work where test were simply made by increasing the order of the power series
by one, the forward stepwise regression scans over all possible next
terms in the power series up to a user defined 11$^{th}$ order.     The results shown above are in the algebraic 
notation of R and the x term is $Q^2$.   Starting from a linear function, different combinations of higher order functions
are tried by adding and removing terms, in order to find the best performing predictive model.   The final result
of the fit is illustrated in Fig.~\ref{R-stepwise} and code to preform the fits can be found on github~\cite{Higinbotham:2016github}.

In the context of assigning meaning to the extracted moments, the stepwise regression result is laughable.
On the other hand, considering these fits as only descriptive approximations valid over the range of the data,
the result is a function with uncertainty that describes the data well.

It is the fact that there is nearly an infinite set of functions that can fit the data within the 
error bands that makes the Monte Carlo modeling and the use of model selection criteria so important; otherwise
one is simply doing a deep search in order to find a particular solution. 

\section{Conformal Mapping}
\label{appendix-mapping}

It has been argued that the use of conformal mapping can improve the situation; but while some authors of this
technique confine themselves to low $Q^2$ and model constrained parameters, others seem to use the
mapping to obfuscate clear over-fitting.    
For example, in Table~1 of~\cite{Hill:2010yb}
the authors do various fits up to fifth order with no clear model selection technique in mind.
By simply using the selection criteria provided herein, AIC or BIC, one finds that
in general the two parameter fits are sufficient and the data used therein 
give 0.86(2)fm for the radius.   By model selection criteria, all the fits beyond second 
order are simply over-fitting the data.   

On the other hand, the idea of conformal mapping is very powerful.   From a pure mathematical point
of view, it can take a function, like $G_E(Q^2)$ that goes from zero in infinity and map it onto a
finite range where one could use orthogonal polynomial to describe data.   It can also be used to
avoid the effects of poles when doing polynomial regressions, though it is worth noting
that rational functions also have that property~\cite{NR}.

\begin{figure}
\includegraphics[width=\columnwidth]{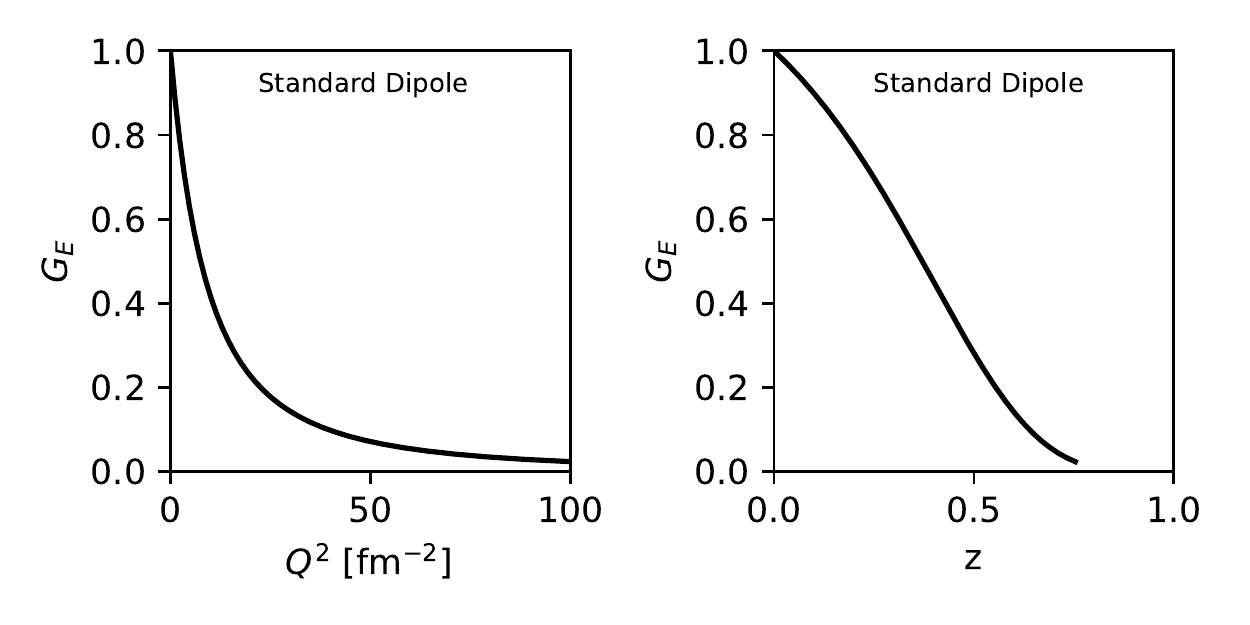}
\includegraphics[width=\columnwidth]{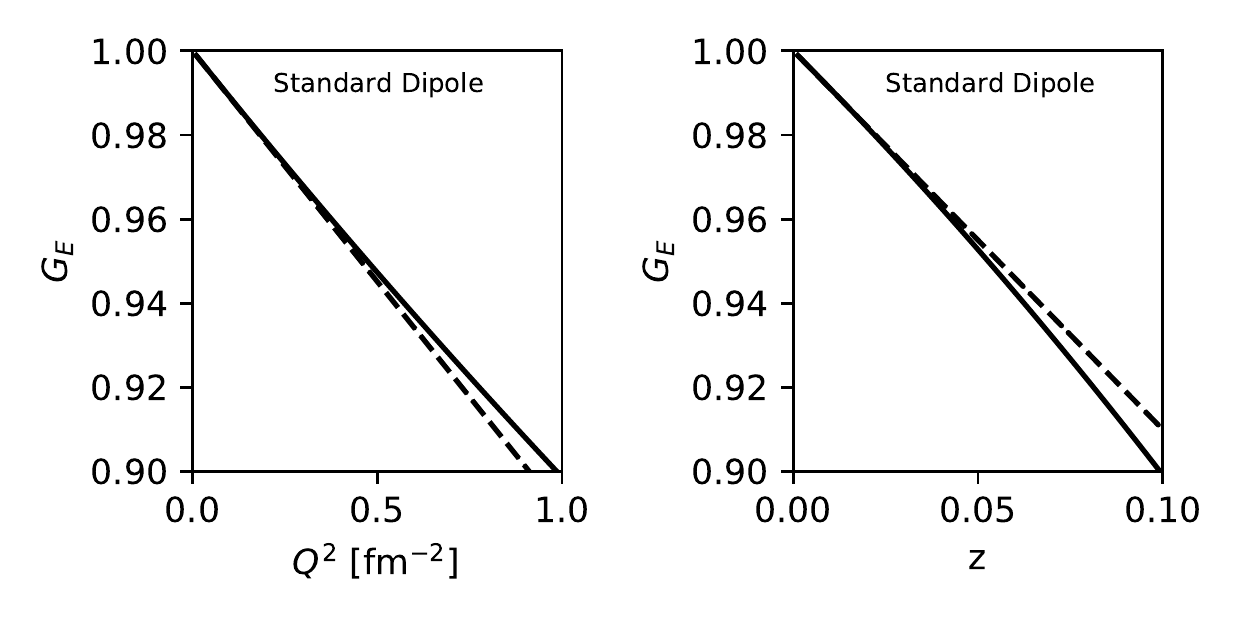}
\caption{Illustration of the result of performing the z transformation on the standard dipole function
shown the solid line. 
For the standard dipole at low $Q^2$, this turns a slightly concave function into a convex function.
The dashed line is simply a straight line to make the concavity of the functions clear. 
As described in the text, this same effect is seen with real $G_E$ data.}
\end{figure}

The transformation from $Q^2$ to the mapped parameter $z$ is done by:

\begin{equation}
z(t,t_{\rm cut},t_0) = \frac{\sqrt{t_{\rm cut} - t} - \sqrt{t_{\rm cut} - t_0} }{ \sqrt{t_{\rm cut} - t} + \sqrt{t_{\rm cut} - t_0}  },  
\end{equation}
where $t=-Q^2$, $t_{\rm cut} = 4m_\pi^2$, and $t_0$ is a free parameter 
representing the point being map onto $z=0$.   
Herein we have used $t_0$ = 0.  
It is argued without proof that ``the curvature is smaller in the
z variable than in the $Q^2$ variable''~\cite{Hill:2010yb}, yet by taking 
exactly the same data as before and fitting with stepwise regression
one gets the following result:

\begin{Verbatim}[fontsize=\footnotesize]

Start:  AIC=62.02
data$y ~ data$x

               Df Sum of Sq    RSS    AIC
+ I(data$x^2)   1    36.992 358.23 31.779
+ I(data$x^3)   1    35.577 359.65 33.092
+ I(data$x^4)   1    33.518 361.71 34.993
+ I(data$x^5)   1    31.295 363.93 37.034
+ I(data$x^6)   1    29.136 366.09 39.003
+ I(data$x^7)   1    27.125 368.10 40.827
+ I(data$x^8)   1    25.277 369.95 42.495
+ I(data$x^9)   1    23.582 371.64 44.017
+ I(data$x^10)  1    22.026 373.20 45.408
+ I(data$x^11)  1    20.595 374.63 46.683
<none>                      395.23 62.019

Step:  AIC=31.78
data$y ~ data$x + I(data$x^2)

               Df Sum of Sq    RSS    AIC
<none>                      358.23 31.779
+ I(data$x^7)   1   0.46895 357.77 33.828
+ I(data$x^6)   1   0.46636 357.77 33.830
+ I(data$x^8)   1   0.46557 357.77 33.831
+ I(data$x^9)   1   0.46025 357.77 33.836
+ I(data$x^10)  1   0.45531 357.78 33.840
+ I(data$x^5)   1   0.45194 357.78 33.844
+ I(data$x^11)  1   0.45177 357.78 33.844
+ I(data$x^4)   1   0.41901 357.82 33.874
+ I(data$x^3)   1   0.36310 357.87 33.926

Weighted Residuals:
     Min       1Q   Median       3Q      Max
-3.07706 -0.75955 -0.09171  0.67092  3.04833

Coefficients:
              Estimate Std. Error  t value Pr(>|t|)
(Intercept)  0.9992836  0.0005757 1735.683  < 2e-16 
data$x      -0.9741340  0.0256596  -37.964  < 2e-16 
I(data$x^2) -1.5340681  0.2627947   -5.838 1.27e-08 

Residual standard error: 1.042 on 330 degrees of freedom
Multiple R-squared:  0.9932,  Adjusted R-squared:  0.9932
F-statistic: 2.41e+04 on 2 and 330 DF, p-value: < 2.2e-16

\end{Verbatim}

\begin{figure}[htb]
\includegraphics[width=\columnwidth]{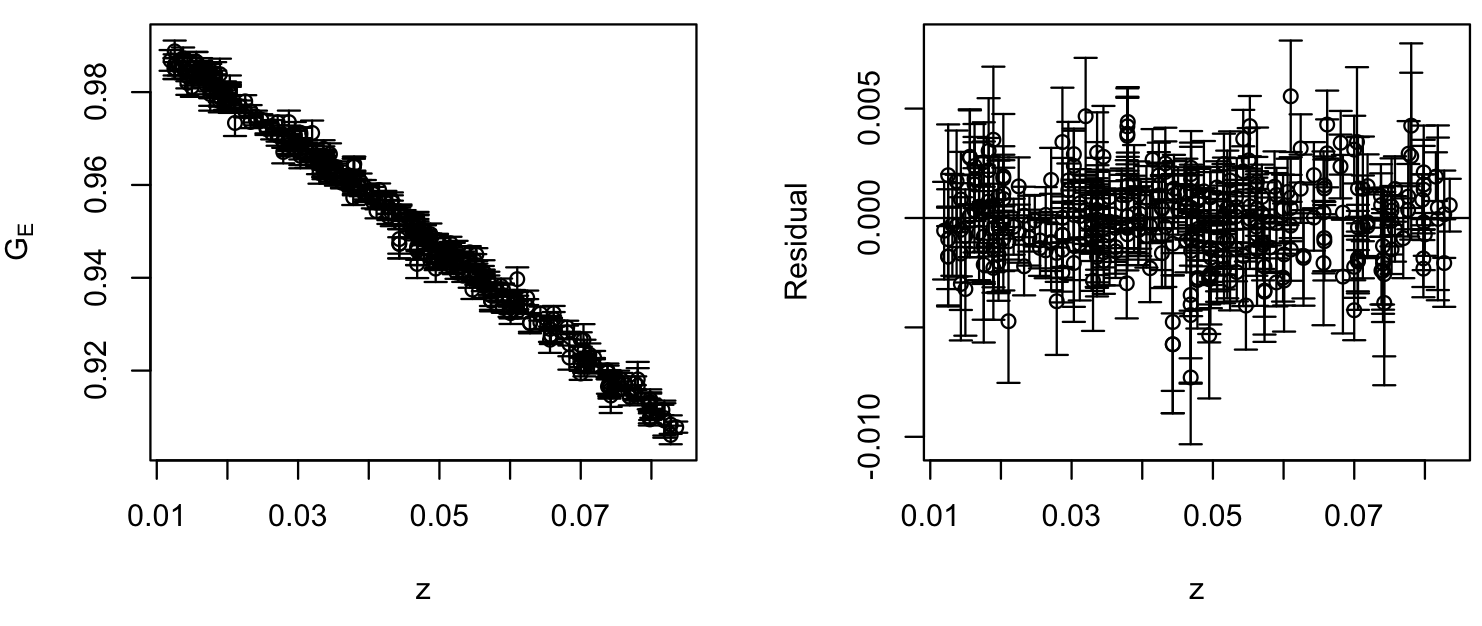}
\caption{The z-transformed $G_E$ data shown along with a residual of the best fit.}
\end{figure}

Thus, the stepwise regression shows that the transformation
has increased the curvature of the data and turned the slightly
convex data in $Q^2$ to clearly concave data in $z$.
Assuming that a quadratic can be used to make a reasonable extrapolation, one can
use this R regression result to extract a radius using the formula:
\begin{equation}
r_p = \frac{\sqrt{-6 (data\$x)}}{4 m_{\pi}}.
\end{equation}
Thus, one finds 0.85(1)~fm,  a result again consistent with the muonic Lamb shift result,
as well as the other results shown in this work.
These ideas can be taken further and one could use ideas such as Gaussian process 
regression to calculate many possible extrapolating paths and then assign relative 
probabilities~\cite{Rasmussen:2005} as recently done by Zhou {\it{ et al.}}~\cite{Zhou:2018bon}.

\section{Semi-analytical Calculations}\label{SemiAnalytical}

In this section we demonstrate a semi-analytical procedure to calculate the bias and variance induced by the noise in our estimation of the slope at zero. This framework can be used to reproduce the graph shown in Fig.~\ref{mc-results} and Fig.~\ref{mihaplot} without the repeated Monte Carlo procedure of creating many noisy data sets and sampling from the fitted functions. The virtue of this method is that it can give us explicitly the influence each data point is having on our bias and variance and, thus, can be used to identify the points that maximize the information gained.

Before proceeding we must realize that the bias in our estimation of the slope has two sources. The first one is due to the noise in the data and will increase as the noise increases. The second is due to the possible fact that we are not using the ``true'' function to fit the data, as was shown in Table~\ref{simpleVSperfect} when the line had a clear bias even when the noise was very small. We therefore split our bias as:

\begin{equation}\label{biases}
 bias=bias_{\sigma}+bias_{0},
\end{equation}
where $bias_{\sigma}$ denotes the bias that scales with the noise, while $bias_0$ will always be present even when $\sigma \rightarrow 0$, where $\sigma$ represents the size of the noise.

In this section we will show how to calculate the first source, while the second can only be calculated if we know the ``true'' value of the slope. Let us define our quantity of interest, the slope at zero, as $m(X,Y,\epsilon)$, where $X$ and $Y$ are the lists of $n$ data points and $\epsilon$ is a particular realization of the noise, which we assume to be Gaussian with mean zero and deviation $\sigma$. Here $m(X,Y,\epsilon)$ can be a closed expression for the slope given the data, as is the case for linear models, or it can be treated as a numerical routine that returns the slope when fitting a non linear model, like the Dipole.

\subsection{Noise Bias}

We want to find the average value of $m(X,Y,\epsilon)$, $\braket{m(X,Y)}$, once all the possible realizations of the noise have been taken into account, weighted correctly by their Gaussian distribution. The result is given by the integral:

\begin{equation}\label{Integral}
  \braket{m(X,Y)}\equiv \int P(\mathbf{\epsilon})\ m(X,Y,\mathbf{\epsilon}) \ d\epsilon,
\end{equation}
where $d\epsilon=d\epsilon_1 d\epsilon_2 ... \ d\epsilon_n$, and $ P(\mathbf{\epsilon})$ denotes the Gaussian probability distribution for the noise given by:

\begin{equation}\label{Prob}
P(\mathbf{\epsilon})=\left(\frac{1}{(\sqrt{2\pi \sigma^2})}\right)^n e^{(\epsilon_1 ^2 + \epsilon_2 ^2 + ... \ \epsilon_n ^2 )/2\sigma^2}.
\end{equation}

The integral in Eq.~\ref{Integral} is effectively taking into account all possible noise realizations, weighted by their corresponding probabilities. 

Since in most cases we do not have an available expression for $m(X,Y,\epsilon)$, we can expand it using a multivariate Taylor expansion and evaluate each term directly under the integral sign in Eq.~\ref{Integral}. The first three terms in this expansion are:

\begin{equation}\label{approx}
m(X,Y,\epsilon)\approx m_0+\left[\overrightarrow{\nabla}_\epsilon m\right]_0 \cdot \overrightarrow{\epsilon} + \frac{1}{2} \overrightarrow{\epsilon}^T \cdot \left[\mathcal{H}_{m}\right]_0\cdot \overrightarrow{\epsilon},
\end{equation}
where $m_0=m(X,Y,\epsilon=0)$ is the value of the slope when the noise is zero. We denote $\left[\overrightarrow{\nabla}_\epsilon m\right]_0$ the gradient of $m$ taking the errors $\epsilon$ as variables, $\overrightarrow{\nabla}_\epsilon m \equiv \left(\frac{\partial m}{\partial \epsilon_1}, \frac{\partial m}{\partial \epsilon_2}, \ ... \frac{\partial m}{\partial \epsilon_n}\right)$, evaluated at $\epsilon=0$. Finally, $\left[\mathcal{H}_{m}\right]_0$ denotes the Hessian matrix of $m$, $\mathcal{H}_{m\ i,j}\equiv \frac{\partial^2 m}{\partial \epsilon_i \partial \epsilon_j}$, again taking the $\epsilon$ as variables and evaluating at $\epsilon=0$. 

Equation~\ref{approx} says that for some small realization of the errors $\epsilon_i$, the value of $m$ is approximately its value when $\epsilon=0$, plus a linear correction in $\epsilon$ by the gradient and finally a second order correction proportional to $\epsilon^2$ that involves the Hessian. Both of these quantities, the gradient and the Hessian are evaluated at $\epsilon=0$ and are therefore just a list and a matrix of fixed numbers, respectively. These two groups of numbers can be obtained numerically by taking finite differences on $m$ as an approximation to derivatives.

Once we have expanded our slope function $m$ we proceed to calculate the integral~\ref{Integral}:

\begin{equation}
\begin{aligned}
&\braket{m(X,Y)}\approx \\
&\int \left( m_0+\cancelto{0}{\left[\overrightarrow{\nabla}_\epsilon m\right]_0\cdot \overrightarrow{\epsilon}} + \frac{1}{2} \overrightarrow{\epsilon}^T \cdot \left[\mathcal{H}_m \right]_0\cdot \overrightarrow{\epsilon}\right) P(\mathbf{\epsilon})d\epsilon.
\end{aligned}
\end{equation}

The linear correction, being proportional to $\epsilon$, would integrate to zero. The term $m_0$ does not depend on the noise and since $P(\epsilon)$ integrates to $1$ it would just appear as it is in the final result. The last term with the Hessian is a quadratic form that would look like $\epsilon_1^2 H_{11} + 2\epsilon_1 \epsilon_2 H_{12} +\ ... \ + \epsilon_n^2 H_{nn}$, where the constants $H_{i,j}$ are the $(i,j)$ elements of the Hessian. Since our noise additions are mutually independent ($\epsilon_i$ and $\epsilon_j$ do not correlate) the integration of terms involving mixing of different $\epsilon_i$ will also yield zero. The integration over terms that involve the same $\epsilon_i$ are by definition the variance of the noise, $\sigma^2$. Therefore we have:

\begin{equation}\label{finalhess}
  \begin{aligned}
  \braket{m(X,Y)} &\approx m_0+ \frac{1}{2}\int \left( \overrightarrow{\epsilon}^T \cdot \left[\mathcal{H}_m \right]_0\cdot \overrightarrow{\epsilon}\right) P(\mathbf{\epsilon}) d\epsilon \\
  &= m_0+\frac{1}{2}\sigma^2 \left[   \frac{\partial^2 m}{\partial \epsilon_1^2}+\frac{\partial^2 m}{\partial \epsilon_2^2}+\ ...\ +\frac{\partial^2 m}{\partial \epsilon_n^2}\right]_0.
  \end{aligned}
\end{equation}

This equation shows that there will be a noise-induced bias in the average estimation of our quantity $m$, which grows proportionally to $\sigma^2$. This bias will not be present if the function that estimates $m$ in terms of the data is linear in the observations $Y$ (and therefore linear in the noise $\epsilon$), since second derivatives of $m$ will be zero from the start. This means that, regardless of the size of the noise, linear fits like a straight line, or any polynomial, will not gain a noise-induced bias, and the distributions for $m$ will always center at the zero noise point. 

For non linear fits on the other hand, there will be a bias that will grow quadratically in the noise. This can apply to the actual function that generated the data, like the Dipole in our study. Once the noise gets too high, the true generating function might not be the most reliable fit to the data, bias-wise. Nevertheless, the effect will usually be very small, since it is proportional to $\sigma^2$.

The term accompanying $\sigma^2$, the trace of the Hessian matrix, can be interpreted as the Laplacian of $m$ at zero noise. Consistent with our findings, the Laplacian operator of $f(x)$ at a point $a$ can been related to the rate of change of the average of $f$, $\braket{f}$, at a small sphere centered at $a$ compared to $f(a)$~\cite{styer2015geometrical}.

\subsection{Standard Deviation}

In order to calculate the standard deviation $s=\sqrt{\braket{m^2}-\braket{m}^2}$ of our distribution for $m$ we must compute the expected value of $m^2$, since we already have $\braket{m}$. We can calculate this quantity up to second order in $\sigma$, using expression \ref{approx}:

\begin{equation}
  \begin{aligned}
  &\braket{m(X,Y)^2}\approx \\
  &\int \left[m_0+\left[\overrightarrow{\nabla}_\epsilon m\right]_0\cdot \overrightarrow{\epsilon} +  \frac{1}{2} \overrightarrow{\epsilon}^T \cdot \left[\mathcal{H}_m \right]_0\cdot \overrightarrow{\epsilon}  \right]^2 P(\mathbf{\epsilon})d\epsilon  = \\ &\int \bigg[ m_0^2+\cancelto{0}{2m_0 \left[\overrightarrow{\nabla}_\epsilon m\right]_0\cdot \overrightarrow{\epsilon}} + \left( \left[\overrightarrow{\nabla}_\epsilon m\right]_0\cdot \overrightarrow{\epsilon}\right)^2 + \\ & 2m_0 \frac{1}{2} \overrightarrow{\epsilon}^T \cdot \left[\mathcal{H}_m \right]_0\cdot \overrightarrow{\epsilon} + \mathcal{O} (\epsilon^4) \bigg] P(\mathbf{\epsilon}) d\epsilon .
  \end{aligned}
\end{equation}

The first term does not depend on $\epsilon$ and can go out of the integral. The second term is proportional to $\epsilon$ and integrates to zero. The Hessian term we already know how to integrate and we have neglected its square, which will be order $\epsilon^4$. The third term, the gradient square, reads:

\begin{equation}
\left( \left[\overrightarrow{\nabla}_\epsilon m\right]_0\cdot \overrightarrow{\epsilon}\right)^2= \left( \frac{\partial m}{\partial \epsilon_1}\bigg|_0 \epsilon_1+ \frac{\partial m}{\partial \epsilon_2}\bigg|_0 \epsilon_2+ \ ... +\frac{\partial m}{\partial \epsilon_n}\bigg|_0 \epsilon_n  \right)^2.
\end{equation}

Once we expand the square we will have two types of terms, those of the form $\left(\frac{\partial m}{\partial \epsilon_i}\big|_0 \right)^2 \epsilon_i^2$ and those with mixed $\epsilon$, $(\frac{\partial m}{\partial \epsilon_i} \frac{\partial m}{\partial \epsilon_j})\big|_0 \epsilon_i \epsilon_j$, $i\neq j$. Since our noise components are independent, only the first type of terms will give non zero results under integration: $\sigma^2$. Therefore we have:

\begin{equation}\label{gradexpand}
\begin{aligned}
& \braket{m(X,Y)^2}\approx m_0 + m_0\sigma^2Tr[\mathcal{H}_m]_0 + \\ &\sigma^2 \left[ \left( \frac{\partial m}{\partial \epsilon_1}\right)^2+ \left(\frac{\partial m}{\partial \epsilon_2}\right)^2 \ ... \ +\left( \frac{\partial m}{\partial \epsilon_n}\right) ^2  \right]_0.
\end{aligned}
\end{equation}

The first coefficient accompanying $\sigma^2$ is the trace of the Hessian matrix, which will cancel the same exact term that appears in the square of the mean value for $m$ : $\braket{m}^2 \approx (m_0 + \frac{1}{2}\sigma^2 Tr[\mathcal{H}_m]_0)^2 \approx m_0^2 + m_0\sigma^2 Tr[\mathcal{H}_m]_0  $.

The second coefficient accompanying $\sigma^2$ is the magnitude of the gradient of $m$, evaluated at zero noise. Our result for the standard deviation $s$, if we only keep terms up to first order in $\sigma$ for $\braket{m}^2$ then reads:

\begin{equation}\label{finalgrad}
s=\sqrt{\braket{m^2}-\braket{m}^2} = \sigma \|\overrightarrow{\nabla}_\epsilon m \|_0.
\end{equation}

Therefore, our calculations show that no matter how small the noise in the data, there would be an appreciable, proportional, standard deviation in our estimate of $m$. This is true unless the gradient is zero, meaning that our quantity $m$ is not sensitive to change in the data to first order.

All the calculations presented here can easily be generalized to include correlations between the $\epsilon_i$ and to include different sizes of noise for each point: $\sigma_i$ instead of a single $\sigma$. The only requirement is that the noise is considered Gaussian and the integrals would still be easily solvable. 

It is also possible to extend the Taylor expansion of $m$ in equation ~\ref{approx} to include more orders in the noise $\epsilon$. Since the Taylor expansion will contain some combination of different $\epsilon_i$, the Gaussian integrals can still be performed in closed form, term by term.

\subsection{Maximizing Information Gained}

Some fundamental questions that arises in experimental designs and theory of information are: for a given relation $f(x)=y$ between two quantities, what are the most relevant values of $x$ such that, when $y$ is measured, the relation $f$ is constrained the most? How much new information one gains by adding more points or reducing the statistical uncertainties on the existing ones?  In other words, what are the $(x,y)$ values that, considered as a set, contain the most amount of information regarding $f$, for a given set of experimental uncertainties. 

With respect to the proton radius extraction that we have studied here, a question of interest would be: for a given range in $Q^2$, what specific values $Q^2_i$ give us the most amount of information regarding the slope at zero, or the proton radius.

Using the semi-analytical framework described, we can attempt to give a quantitative answer to this question.  Both Eqs.~\ref{finalhess} and ~\ref{gradexpand} have explicit the contribution each point makes to the noise bias and the standard deviation, respectively. Therefore, we could directly identify how a reduction in uncertainty on each particular point would propagate to our estimate of the radius.

Alternatively, if we are stuck on some level of uncertainty, we could estimate where the $n$ measurements should be made such that, again, our information gain is maximal. In order to test these ideas, we minimized the RMSE defined in Eq.~\ref{RMSEdef} as a function of the location of $n$ $Q^2$ points, and studied how much our uncertainty in the slope at zero is reduced as $n$ increased.

Our $y$ points were again generated by the Dipole function defined in Eq.~\ref{sd} and we studied the fit of a line in the range $ 0.1 - 0.8 $ fm$^{-2}$ and the fit of a parabola in the range $ 0.1 - 1.6 $ fm$^{-2}$, both of which have proven to be optimal in our simulations on section ~\ref{Goldilocks}. 

Since the slope estimate from both models is linear in the observations $y$, we know that there will be no noise-induced bias, and the only source of bias will be a constant term, see Eq.~\ref{biases}. By ``constant'' we mean that it will not grow with the noise, but it would change depending on the points $Q^2_i$ used. 

We seek therefore to minimize:

\begin{equation}\label{RMSEExpanded}
\begin{aligned}
& {\mathrm{RMSE}}^2(X) = bias^2 (X) + s^2 \\ 
& = (m_{true} - m_0(X))^2 +  \sigma^2 \|\overrightarrow{\nabla}_\epsilon m (X) \|_0 ^2,
\end{aligned}
\end{equation}
where $X=(Q^2_1,\ ... \ Q^2_n)$ are the $n$ locations of the ``observed'' points, and we have omitted the $Y$ dependence on our quantities since we are calculating $Y$ directly from our points $X$, and  $m_{true}=-0.1097$ is the fixed value of the slope at zero, in fm$^2$.

Now, the bias quantity ($m_{true} - m_0(X)$) will indeed depend on $X$, and is a number that is usually hidden to us by nature, since we do not know the true value. Nevertheless, this exercise is helpful in showing that some points have more information than others, and we will see how both terms evolve as $n$ grows.

\begin{figure}[htb]
\includegraphics[width=\columnwidth]{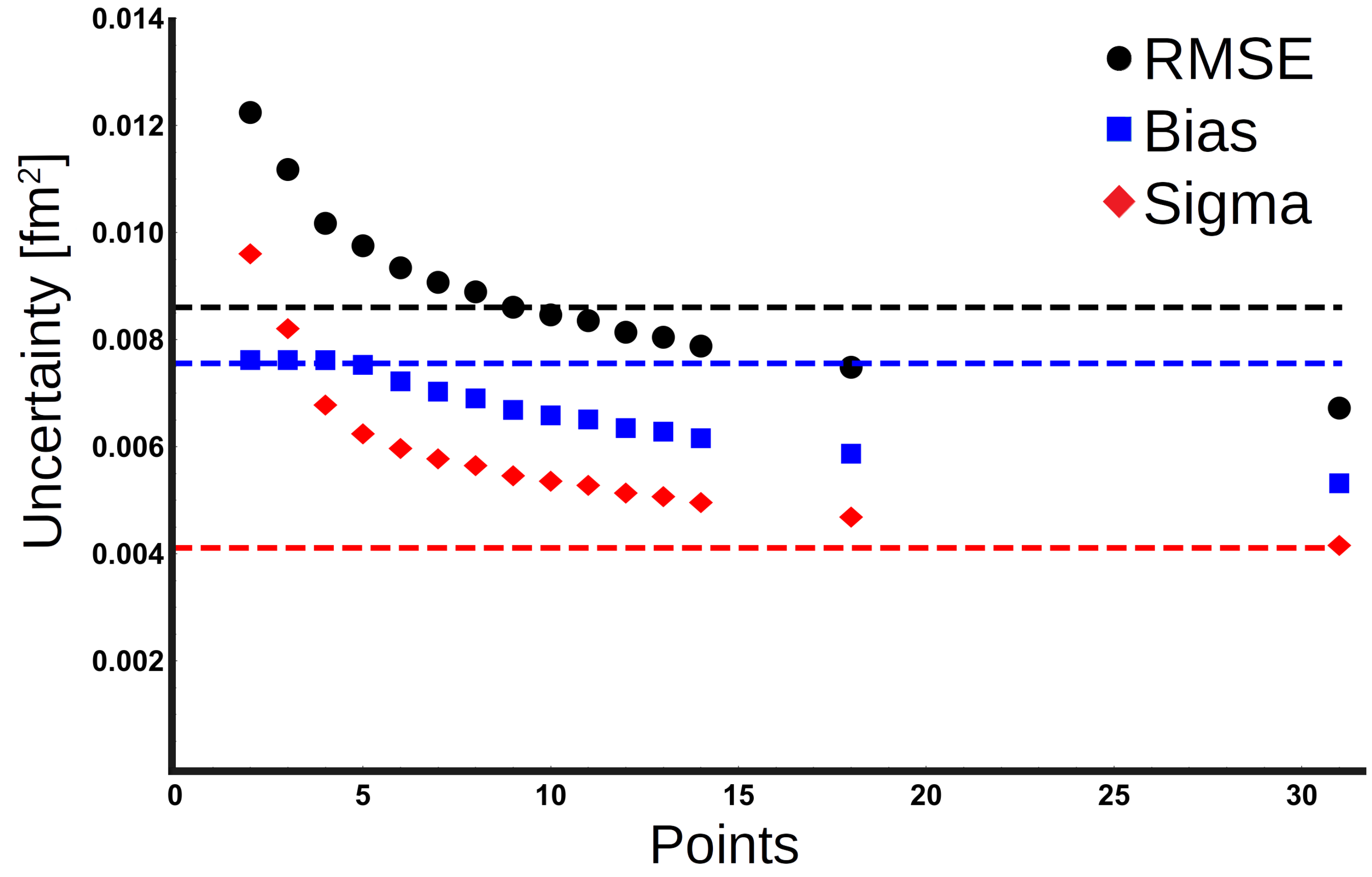}
\caption{
Line RMSE, absolute Bias and Sigma, $s$, for the optimal location of points as a function of the number of points available, in the range 0.1-0.8 fm$^{-2}$. The dashed horizontal lines show the respective values for the configuration of 31 equidistant points showed in Table~\ref{fulltable}.}
\label{GraphLine}
\end{figure}

\begin{figure}[htb]
\includegraphics[width=\columnwidth]{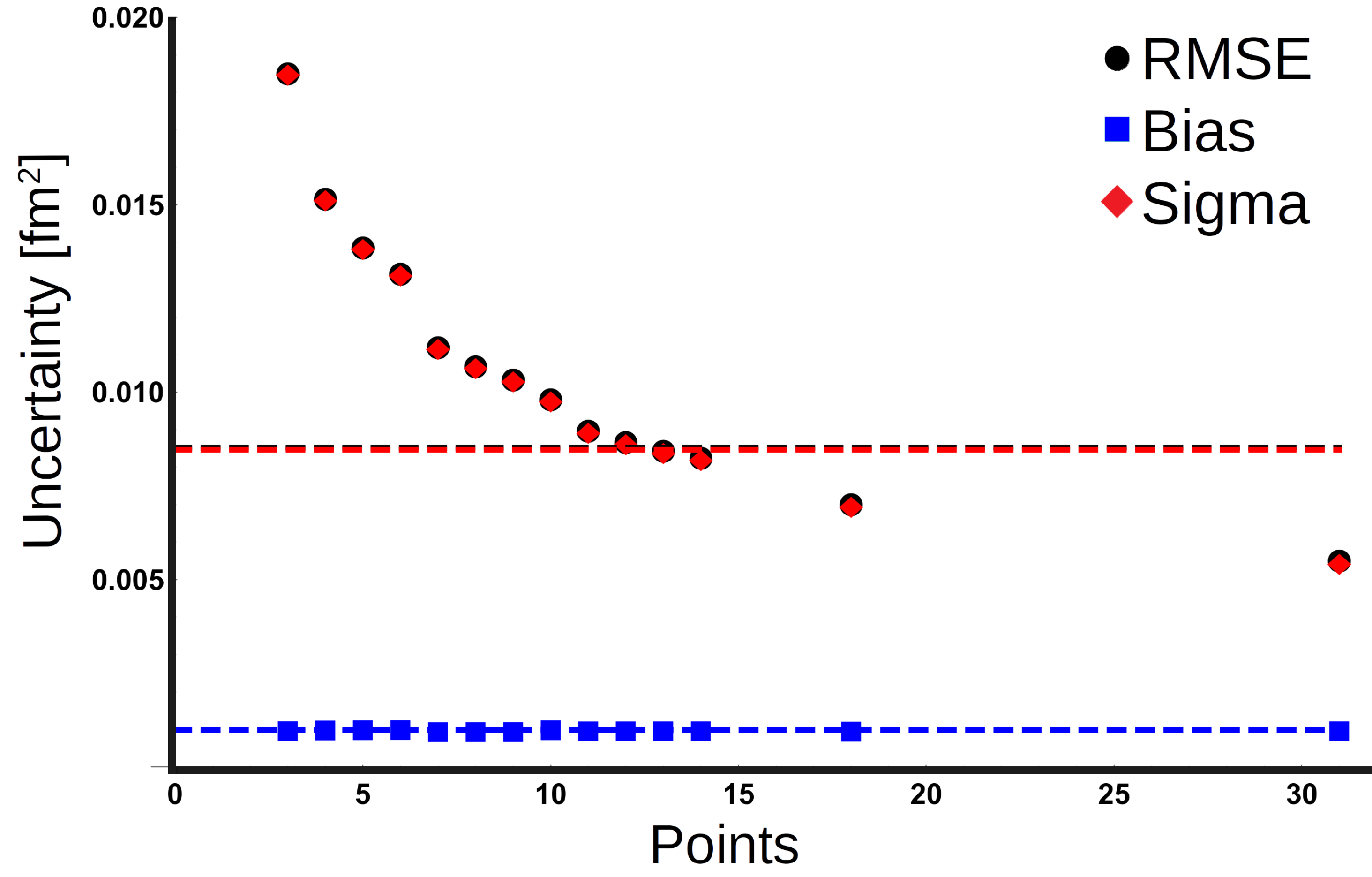}
\caption{
Parabola RMSE, absolute Bias and Sigma, $s$, for the optimal location of points as a function of the number of points available, in the range 0.1-0.1.6 fm$^{-2}$. The dashed horizontal lines show the respective values for the configuration of 31 equidistant points showed in Table~\ref{fulltable}.}
\label{GraphParabola}
\end{figure}

Figures~\ref{GraphLine} and ~\ref{GraphParabola} show the obtained results when the points $Q^2_i$ are not distributed uniformly but are rather moved to positions that, respecting the assigned range, minimize the RMSE given by Eq.~\ref{RMSEExpanded}. It is very interesting to note that, as we expected from our analysis in section~\ref{Goldilocks}, the RMSE of the parabola is primarily driven by the standard deviation contribution, while the line is more balanced, although towards higher bias.

The main result that we can extract from these two graphs is that it takes only 10 points in the case of the line and 13 in the case of the parabola to obtain the same value of RMSE compared to 31 uniformly spaced points, see Table~\ref{equaldatatable}. If we use 31 points but locate them in optimal positions we obtain an RMSE of 0.0068 fm$^2$ for the line and an RMSE of 0.0056 fm$^2$ for the Parabola. The first one represents an improvement of 21\% while the second represents and improvement of 31\%.

As we have already mentioned, calculating (and optimizing with respect to) the bias requires the true value to be known, which is usually not the case in interesting problems. Nevertheless, the standard deviation we have defined is computable only as a function of the observed values and errors, and therefore it is a quantity we can use when facing the real data. As can be seen in Figure~\ref{GraphParabola} the bias of the parabola does not seem to change much for any configuration of points, but the $\sigma$, and therefore the RMSE, can be improved substantially.

\pagebreak
\end{appendix}


%

\end{document}